\documentclass[prx,onecolumn,aps,a4paper,superscriptaddress]{revtex4-1}
\usepackage{graphicx}
\usepackage{graphicx} % support the \includegraphics command and options

\usepackage{color}
\usepackage{subfigure}
\usepackage{braket}
\usepackage{amsmath}
\usepackage{amssymb}
\usepackage{verbatim} % adds environment for commenting out blocks of text & for better verbatim
\def\ii{{\rm i}}  \def\ee{{\rm e}}  
\def\be{\begin{equation}}
\def\ee{\end{equation}}
\def\bea{\begin{eqnarray}}
\def\eea{\end{eqnarray}}

\begin{document}

\title{Designing exotic many-body states of atomic spin and motion in photonic crystals}
\author{Marco T. Manzoni}
\affiliation{ICFO-Institut de Ciencies Fotoniques, The Barcelona Institute of Science and Technology, 08860 Castelldefels (Barcelona), Spain}
\author{Ludwig Mathey}
\affiliation{Zentrum f\"ur Optische Quantentechnologien and Institut f\"ur Laserphysik, Universit\"at Hamburg, 22761 Hamburg, Germany}
\affiliation{The Hamburg Centre for Ultrafast Imaging, Luruper Chaussee 149, Hamburg 22761, Germany}
\author{Darrick E. Chang}
\affiliation{ICFO-Institut de Ciencies Fotoniques, The Barcelona Institute of Science and Technology, 08860 Castelldefels (Barcelona), Spain}

%\date{} % Activate to display a given date or no date (if empty),
         % otherwise the current date is printed

\begin{abstract}
Cold atoms coupled to photonic crystals constitute an exciting platform for exploring quantum many-body physics. Here we investigate the strong coupling between atomic internal (``spin'') degrees of freedom and motion, which arises from spin-dependent forces associated with the exchange of guided photons. We show that this system can realize a remarkable and extreme limit of quantum spin-orbital systems, where both the direct spin exchange between neighboring sites and the kinetic energy of the orbital motion vanish. We find that this previously unexplored system has a rich phase diagram of emergent orders, including spatially dimerized spin-entangled pairs, a fluid of composite particles comprised of joint spin-phonon excitations, phonon-induced N\'eel ordering, and a fractional magnetization plateau associated with trimer formation.
\end{abstract}
\maketitle

\section{Introduction}
Rich phenomena in condensed matter arise when quantum spin systems couple to phonons or orbital degrees of freedom of the underlying crystal lattice. Perhaps the most famous example is the spin-Peierls model \cite{peierls,pytte,cross,bursill}, wherein the spin interaction leads to a lattice instability resulting in a ground state of singlet pairs and a bond-ordered density wave. Motivated by this emergence of new physics, it is tempting then to consider the most extreme limit of coupling between spin and motion -- where the spin-carrying particles are completely free to move, and the spin-dependent forces thus dictate the properties of the emergent spatial order. Such a material would constitute a novel ``quantum crystal'' that has not existed before, in which the emergent spatial patterns and spin properties are intricately locked together, and where driving one would automatically affect the properties of the other.

Here, we propose a route toward realizing such a material, using a promising new experimental platform consisting of cold atoms coupled to photonic crystal structures. Photonic crystals \cite{PC} are periodic dielectric structures in which the propagation of light can differ significantly from uniform media. A unique feature of photonic crystals is the appearance of photonic band gaps, where strong interference induced by the periodicity yields a complete absence of propagating modes within some bandwidth. Nominally, an excited atom whose transition frequency resides in the gap would not be able to spontaneously emit; instead, it has been shown that an atom-photon bound state can form, in which the atom becomes dressed by a localized photonic cloud \cite{john_1,kurizki,john_2,douglas,gonzalez}. The tight spatial confinement associated with this photon yields large dispersive forces on proximal atoms that depend on the atomic internal ``spin'' states, which thus realizes the required spin-dependent forces to possibly observe the phenomena described above.

In this paper, we first describe the key features of atoms coupled to photonic crystals. Out of the vast possibilities of engineering choices, we propose a realistic setup that enables the spin-dependent forces to exceed the external lattice potential. This is achieved by trapping atoms in a weak one-dimensional external potential, and controlling the interaction to be of short range. As a first step towards understanding the emergent orders of this system, we first treat the motion of the atoms classically and their spins quantum mechanically. We find an effect reminiscent of the spin-Peierls transition, in which the atoms spatially dimerize and realize a high degree of entanglement within each dimer. We then proceed to a fully quantum model. Using density matrix renormalization group (DMRG), we find a rich variety of quantum phases beyond the spin-Peierls state, such as a state where spin and phonon excitations form composite particles, phonon-induced N\'eel ordering, and spatial trimers associated with magnetization plateaus. This work therefore not only proposes the creation of exotic correlated spin-orbital quantum matter, but the considerations developed here will similarly apply to all hybrid systems of atoms and photonic crystals.

\section{Atoms coupled to photonic crystal waveguides}

\begin{figure}[t]
	\setlength{\unitlength}{1cm}
	\begin{picture}(18,8.2)
	\put(0.2,0){\includegraphics[height=82mm,angle=0,clip]{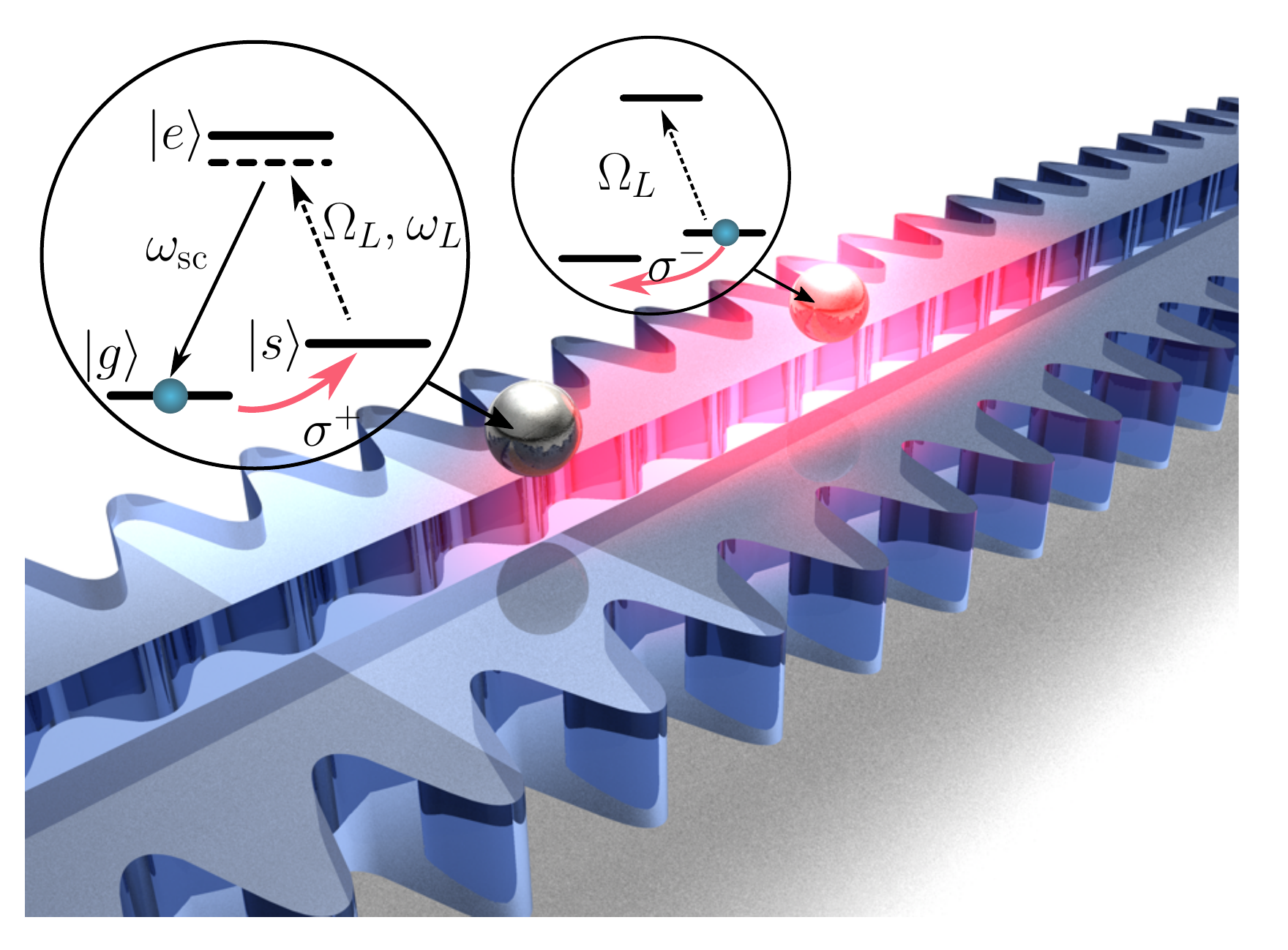}}
	\put(12,1){\includegraphics[height=60mm,angle=0,clip]{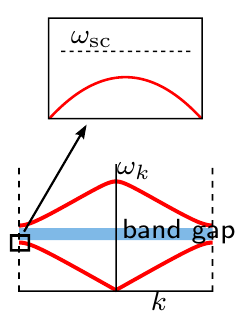}}
	\put(0,8.2){(a)}
	\put(12,8.2){(b)}
	\end{picture}
\caption{\textbf{Atoms interacting with a photonic crystal waveguide.} (a) Schematic 3D rendering of the alligator photonic crystal waveguide (APCW) \cite{yu} with two atoms trapped. The atomic transition $\ket{s}-\ket{e}$ is globally driven by an external laser with Rabi frequency $\Omega_L$. In principle, an atom originally in $\ket{s}$ can Raman scatter a laser photon and flip to state $\ket{g}$. However, when the frequency of the scattered photon $\omega_\mathrm{sc}$ lies within a bandgap (see Fig.~\ref{fig_1}b), this photon becomes bound around the atom (illustrated by the pink cloud). It can be subsequently absorbed by another atom initially in state $\ket{g}$, resulting in a flip to state $\ket{s}$. (b) Illustration of the dispersion relation (frequency $\omega_k$ versus Bloch wavevector $k$) of the guided modes. The scattered photon frequency $\omega_{sc}$ is aligned to a bandgap where no guided modes exist.
}
\label{fig_1}
\end{figure}

Photonic crystals \cite{PC} are periodic dielectric structures in which the propagation of light can differ significantly from uniform media (Fig.~\ref{fig_1}a). The dispersion relation in such structures consists in general of different bands, between which can appear bandgaps -- frequency regions in which the light cannot propagate inside the crystal (Fig.~\ref{fig_1}b).
Particularly rich phenomena are predicted to arise when an atomic transition is driven by a laser at a frequency within the bandgap. A specific example is illustrated in Fig.~\ref{fig_1}a, where two identical atoms are coupled to an ``alligator'' photonic crystal waveguide (APCW), which consists of two separate waveguides whose modes hybridize with one another. Atoms have recently been coupled to such a structure in experiments described in Refs. \cite{yu,goban_1,goban_2,hood}. Here, the atoms are assumed to have three relevant electronic levels. The transition between ground state $\ket{s}$ and excited state $\ket{e}$ is globally driven by an external laser with frequency $\omega_L$ and Rabi frequency $\Omega_L$, while the transition between ground state $\ket{g}$ and $\ket{e}$ is coupled to the guided modes of the waveguide. In principle, an atom in state $\ket{s}$ could Raman scatter a pump photon into the waveguide and flip to state $\ket{g}$. However, when the frequency $\omega_\mathrm{sc}=\omega_L+\omega_{sg}$ of that scattered photon lies within the bandgap (see inset of Fig.~\ref{fig_1}b), it is unable to propagate and instead forms a bound state around the atomic position. A second atom nearby in state $\ket{g}$ can subsequently absorb that photon, resulting in an effective spin flip interaction between the two atoms \cite{douglas}. The effective spin Hamiltonian, generalized to many atoms, takes the form
 \begin{equation}
\label{hamiltonian_1}
H^\mathrm{int} = \frac{J}{2}\,\sum_{i,j}\,f(x_i,x_j)(\sigma^{+}_i\sigma^{-}_{j} + \text{h.c}),
\end{equation}
with $f(x_i,x_j) = e^{-|x_i-x_j|/L}\,E(x_i)E(x_j)$. $\sigma^{-}$ denotes the spin lowering operator from $\ket{s}$ to $\ket{g}$, and conversely for $\sigma^{+}$. $J$ and $L$ are the strength and characteristic length of the interaction respectively, which are tunable through the laser parameters $\omega_{L},\Omega_L$ \cite{douglas}. $E(x_i) = \cos kx_i$ is the Bloch function associated with the electric field at position $x_i$. Here, we will assume that the atoms are tightly trapped in the transverse direction, such that the position along $x$ is the only dynamical variable. Absent any motional effects (i.e., if $f$ is constant), Eq. \eqref{hamiltonian_1} corresponds to the ``XX'' spin model in 1D \cite{lieb}.

The possibility to tune $J$, $L$ and even the type of spin interaction makes the atom-photonic crystal interface a promising candidate for the simulation of many-body spin models with long-range interaction, when atoms are trapped at fixed positions. Going beyond previous proposals involving atoms to simulate spin models \cite{porras,richerme,jurcevic,micheli,gorshkov,taie,trotzky}, here we propose that a qualitatively new and rich set of phenomena can arise by considering the motion as well. In particular, if one treats the position variables $x_i$ as dynamical degrees of freedom, the Hamiltonian in Eq. \eqref{hamiltonian_1} should be regarded as a spin-dependent potential, wherein the forces experienced by the atoms can depend on the spin correlations. As these forces originate from the dispersive forces associated with photons confined to the nanoscale, their magnitude can be comparable to or much larger than those associated with conventional optical trapping forces \cite{douglas}, implying that the resulting physics can become prominent in such systems. Here we will investigate some of the possibilities to achieve novel strongly correlated phases of spin and motion. In emphasizing the role of spin correlations on motion, we also greatly extend previous ideas involving self-organization of atoms in cavities or waveguides due to optical forces, where the atoms are treated essentially as classical dielectric particles with no internal degrees of freedom \cite{CCK13,BAU10,KLI15,BLA03,ASB05,DOM02}.

\section{Classical motion}

\begin{figure}[t]
	\setlength{\unitlength}{1cm}
	\begin{picture}(18,9)
	\put(0.5,5){\includegraphics[width=165mm,angle=0,clip]{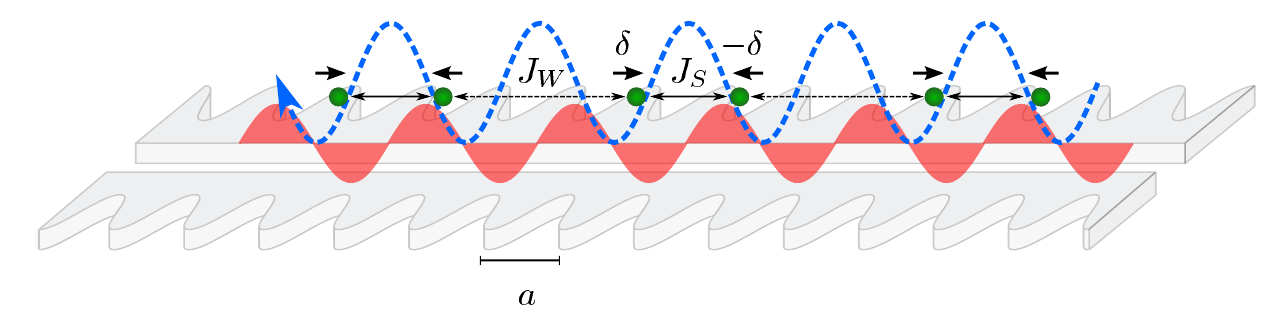}}
	\put(0.3,0){\includegraphics[height=52mm,angle=0,clip]{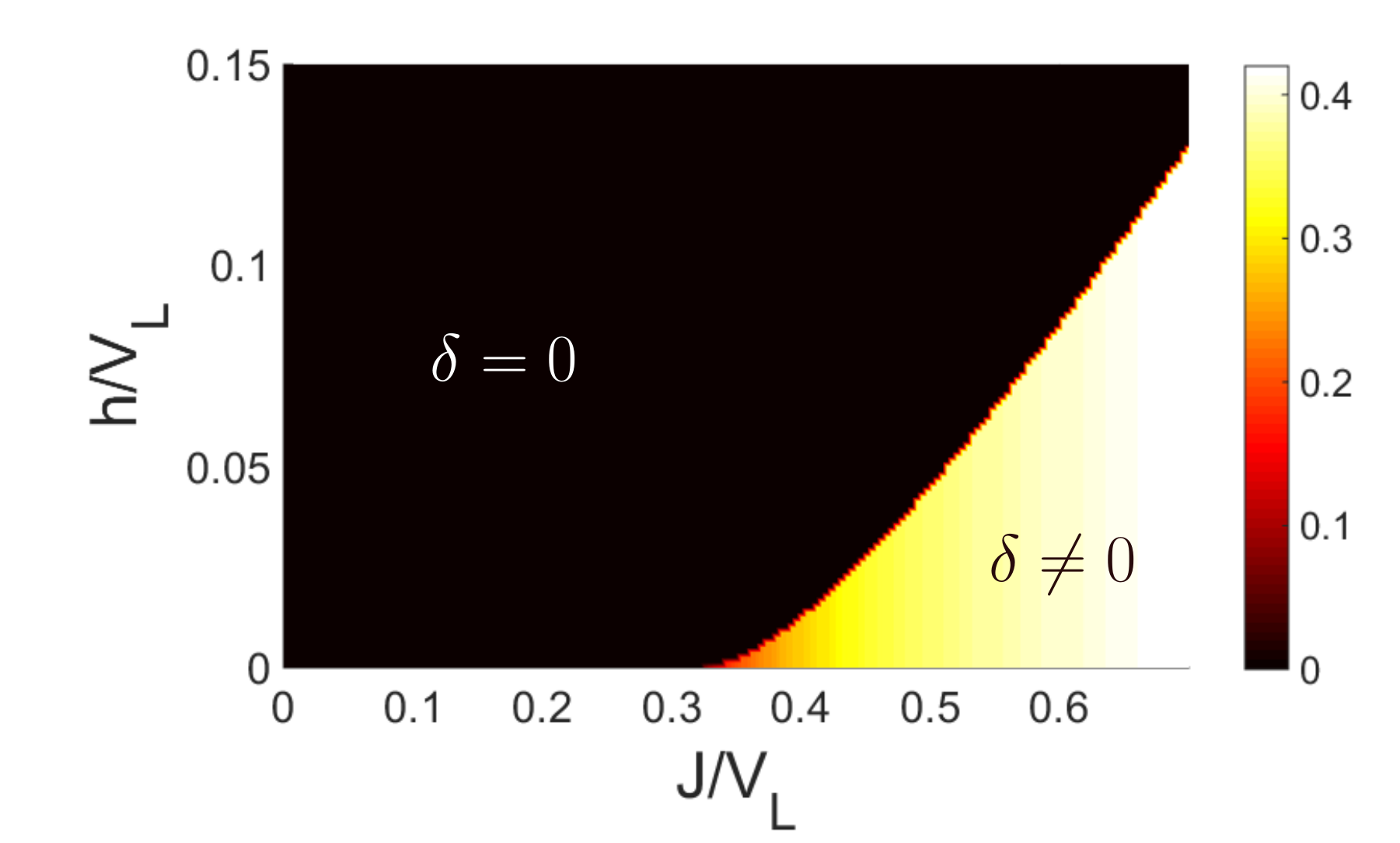}}
	\put(9.5,0.2){\includegraphics[height=49mm,angle=0,clip]{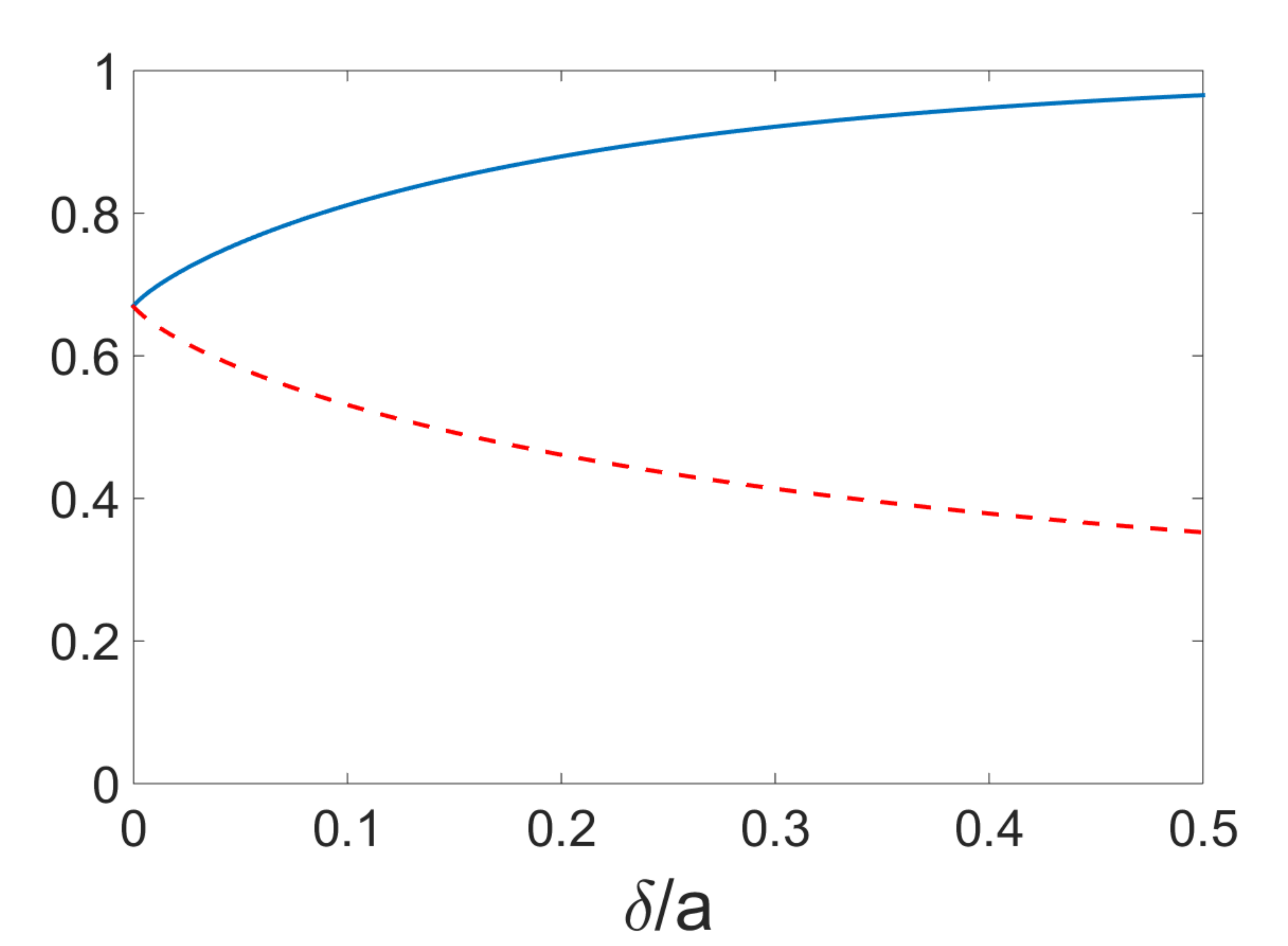}}
	\put(0,9){(a)}
	\put(0,4.5){(b)}
	\put(9.3,4.5){(c)}
	\end{picture}
\caption{\textbf{Spin-motion coupling in the limit of classical motion.} (a) Schematic 1D representation of the model, with atoms (green) trapped in an external potential (blue). The photonic-crystal mediated interaction is modulated by the standing wave of the Bloch modes (red),while the external potential creates trapping sites centred around the nodes. The arrows represent the displacement from the trapping sites to a dimerized configuration. (b) Spatial dimerization $\delta$ (in units of $a$), as function of the interaction strength $J$ and the magnetic field $h$ (in units of $V_L$). (c) Triplet fraction of the reduced density matrix for two atoms within a dimer ($T_S$, blue solid curve), and consecutive atoms in different dimers ($T_W$, red dashed), as a function of dimerization $\delta$, at zero magnetic field ($h=0$).}
\label{fig_2}
\end{figure}
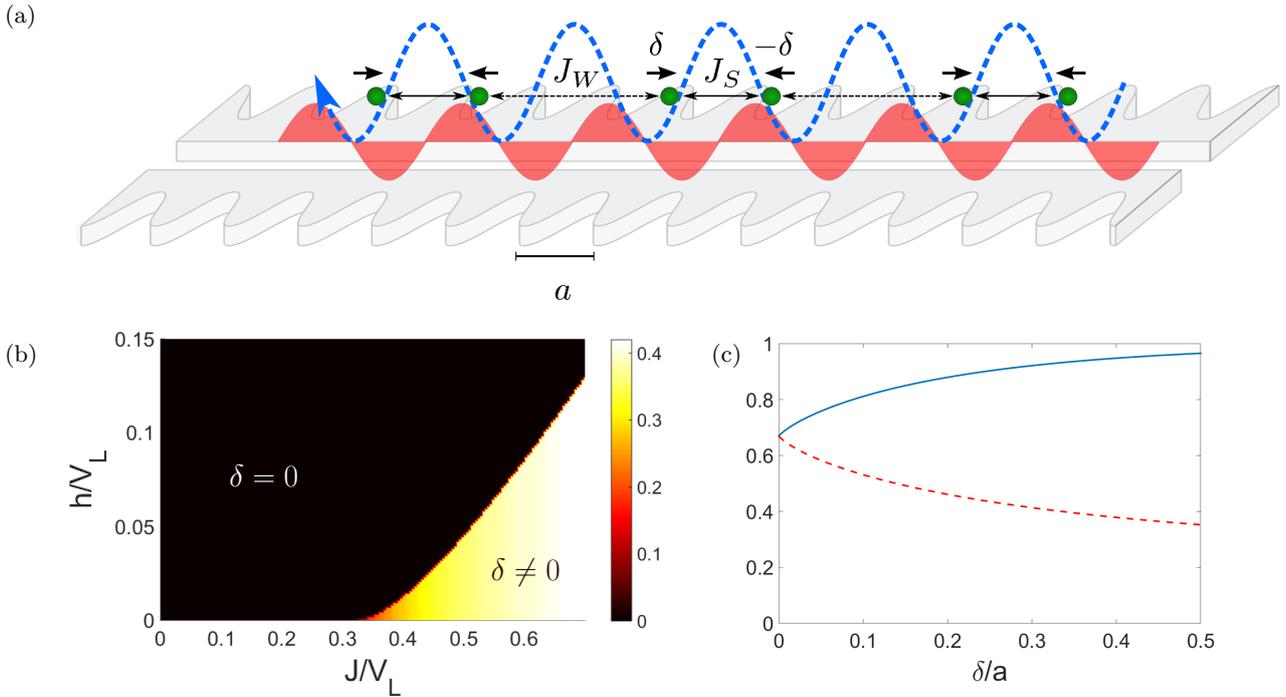

We propose a realistic experimental setup, which highlights the interplay of spin and motion. Atoms interact via the Hamiltonian of Eq.~\eqref{hamiltonian_1}, and are separately trapped by an external, spin-independent optical lattice $H^\mathrm{trap}= V_L\sum_{i} \sin^{2} k_\mathrm{tr} x_i$ (this could originate from optical fields in another guided band far from the atomic resonance). Peculiarly, this lattice traps atoms at the \textit{nodes} of the Bloch function, and thus nominally hides the atoms from the APCW interaction. Despite not being a fundamental requirement to see spin-motion coupling, we assume that the trapping wavelength is such that atoms are localized around even nodes of the Bloch wave functions, i.e. $k_\mathrm{tr} = k/2 = \pi/(2a)$, where $a$ is the length of the unit cell of the APCW, as pictured in Fig.~\ref{fig_2}a. It can be readily shown that within our model, trapping atoms at every site would yield a phase transition with discontinuous change in the atomic positions.

We consider the Hamiltonian in the case of one atom per trapping site and an external magnetic field $h$ along $z$:
\begin{equation}
\label{hamiltonian_class}
H = H^\mathrm{trap} + H^\mathrm{int} + H^\mathrm{magn} = \frac{V_L}{2}\sum_i\,\sin^2 k\delta_i/2 + \frac{J}{2}\sum_{i,j}\,f(x_i,x_j)\,(\sigma^+_i\sigma^-_{j} + \text{h.c}) + h\sum_i\,\sigma^z_i,
\end{equation}
where $\delta_i$ denotes the displacement of atom $i$ from the bottom of its external well. In the present section we treat the atomic position classically, while investigating the case of quantum motion in the next section. We assume that the coupling strength $J$ is positive.

To study the many-body ground state of Hamiltonian \eqref{hamiltonian_class} without any assumption about the spatial configuration is very difficult. Furthermore, for $L/a \gg 1$ the long-range character of the interaction makes the spin model difficult, even for fixed positions. As a consequence, we restrict our attention to the case $L \sim a$, for which we can make a nearest-neighbor approximation.
We can get an intuition of the possible ground state configuration of a system of many atoms by considering how just two atoms in neighboring sites interact. If the atoms remain at the bottom of their trapping wells, the function $f(x_1,x_2)=0$ as these positions coincide with nodes of the Bloch functions. However, the APCW interaction energy would be negative, if the two atoms were to form a triplet state, $\ket{T} = (\ket{sg}+\ket{gs})/\sqrt{2}$ (or a singlet for $J<0$), and simultaneously displace toward each other to form a spatial dimer. Such a process would become energetically favourable overall for a certain ratio of $J/V_L$. 
Motivated by this simple case we make an ansatz that the spatial configuration of the many-body ground state consists of dimerized pairs. In particular, we assume that $x_i = 2ia+(-1)^i\delta$, where $\delta$ represents the displacement from the trap center, as pictured in Fig.~\ref{fig_2}a. This is reminiscent of the lattice instability that creates spatial entangled dimers in the spin-Peierls model \cite{peierls}, but with the substantial difference that our system becomes non-interacting in the absence of dimerization (as the atoms are at the nodes). In the following, we treat $\delta$ as a variational parameter and proceed to solve the spin ground state exactly.

The nearest-neighbor spin Hamiltonian can be mapped to a chain of spinless fermions through standard Jordan-Wigner transformation \cite{JW}, with the presence/absence of a fermion on a site corresponding to spin up/down, respectively. Because of the staggered spatial configuration, it is natural to define a unit cell $j$ consisting of a pair of dimerized atoms (labelled L,R). Two different spin couplings $J_{S,W}(\delta)=J\,\sin^2 k\delta\,e^{-(2a\mp2\delta)/L}$ then characterize the interaction between atoms within the same dimer, and between consecutive atoms $R,L$ in neighboring dimers, respectively (see Fig.~\ref{fig_2}a). The Hamiltonian then reads
\begin{equation}
\label{eq:hamiltonian_class_2}
H(\delta) = \frac{NV_L}{2}\,\sin^2 k\delta/2 - \sum_{j} J_S(\delta)\,(c^\dagger_{L,j}c_{R,j} + \text{h.c})+ J_W(\delta)\,(c^\dagger_{R,j}c_{L,j+1} + \text{h.c}) + 2h(c^\dagger_{L,j}c_{L,j} + c^\dagger_{R,j}c_{R,j} -1),
\end{equation}
where $c_{(L,R),j}$ are fermion annihilation operators for site $j$. Just as in the standard Jordan-Wigner transformation, this two-spin per-site Hamiltonian can be exactly or numerically diagonalized by moving to Fourier space, whose details we describe in Appendix A. 

By minimizing the ground-state energy of $H(\delta)$ with respect to $\delta$ we find the optimal spatial configuration (within the ansatz). In Fig.~\ref{fig_2}b we plot the resulting value of $\delta$ as function of the interaction strength $J$ and of the magnetic field $h$ (in units of $V_L$). In the $J-h$ plane one can clearly distinguish a critical value of the spin interaction strength, $J_\mathrm{crit}(h)$, above which a phase transition occurs from a non-interacting to a dimerized state. The increase in spin entanglement with dimerization can be quantified by taking the two-particle reduced density matrix $\rho_{2S}$ of atoms within a dimer, and calculating its overlap with the triplet state, $T_S(\delta)= \braket{T|\rho_{2S}|T}$. We plot $T_S(\delta)$ in Fig.~\ref{fig_2}c for zero magnetic field. For $\delta = 0$ this quantity tends to the value in the conventional XX spin model, $T_S(0) = (1/2+1/\pi)^2 \approx 0.67$, while for large values of $\delta$ and small $L$ it tends to 1. Similarly, defining an analogous quantity $T_W(\delta)$ between consecutive atoms in neighboring dimers, we find a decrease in correlation with increasing dimerization.

\section{Quantum motion}

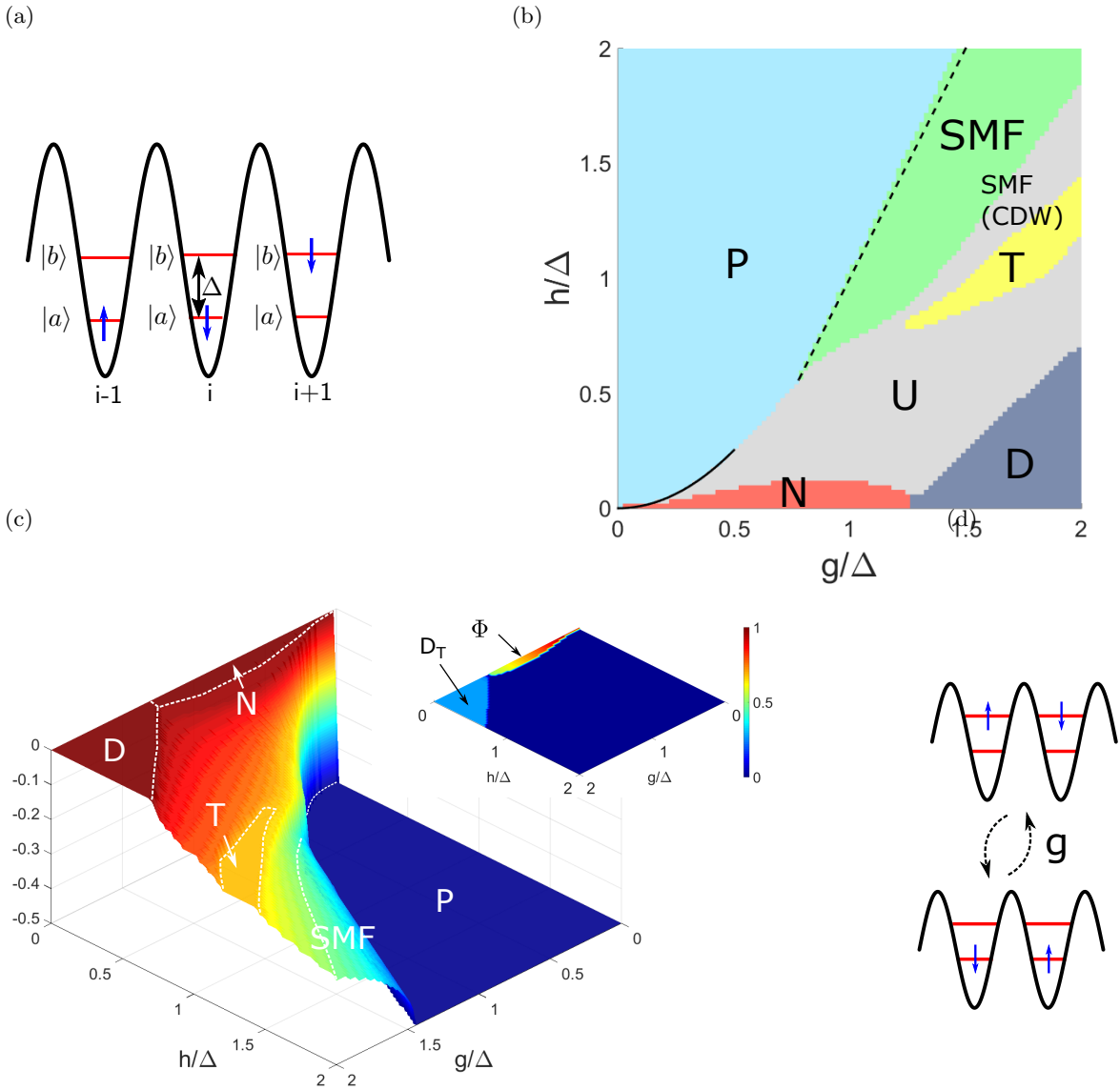
\begin{figure}[ht]
	\setlength{\unitlength}{1cm}
	\begin{picture}(18,15.5)
	\put(0,9.5){\includegraphics[height=40mm,angle=0,clip]{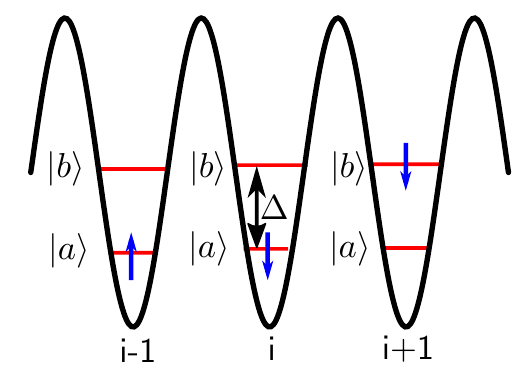}}
	\put(7.2,7.2){\includegraphics[height=78mm,angle=0,clip]{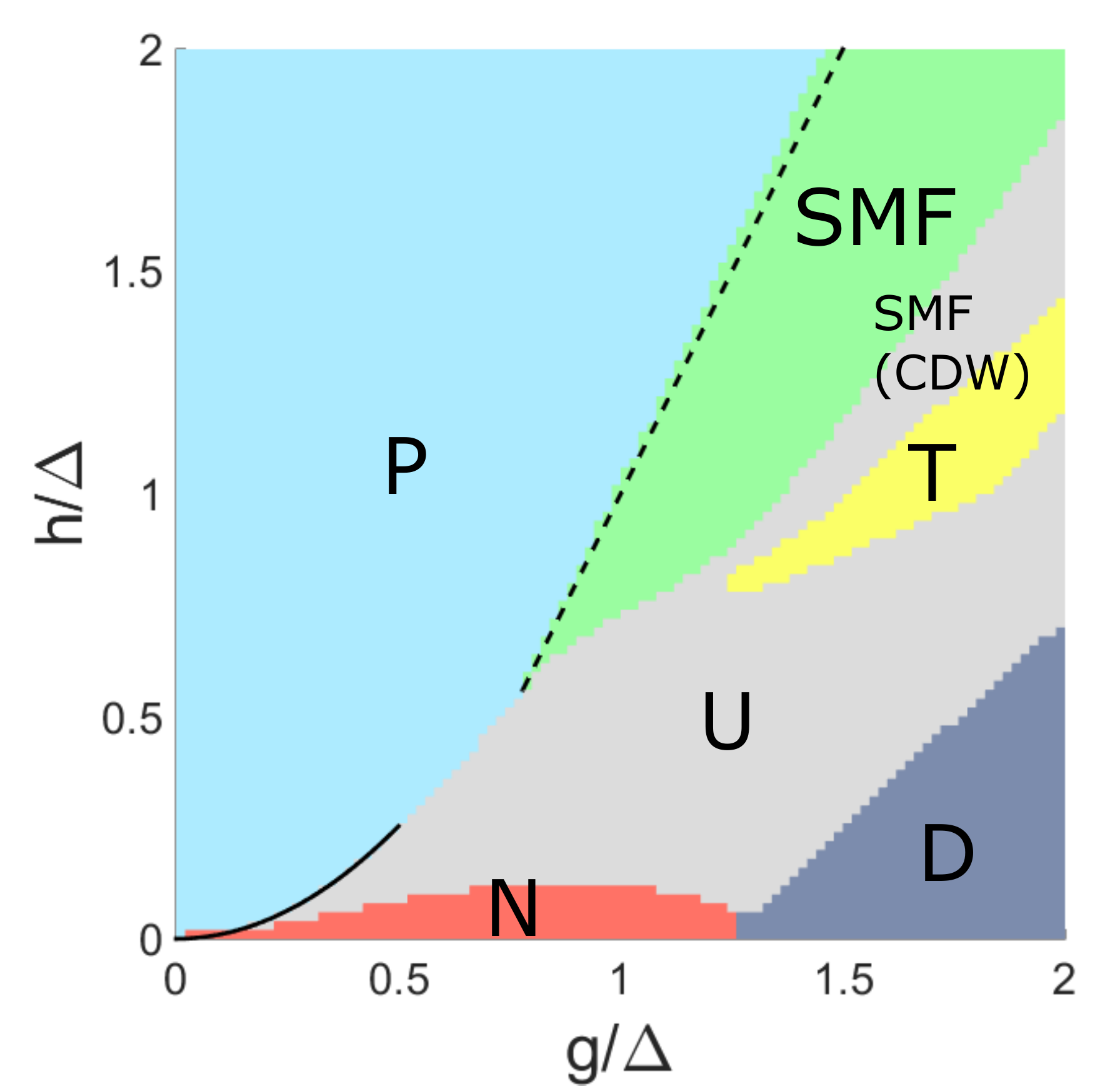}}
	\put(0,0){\includegraphics[height=70mm,angle=0,clip]{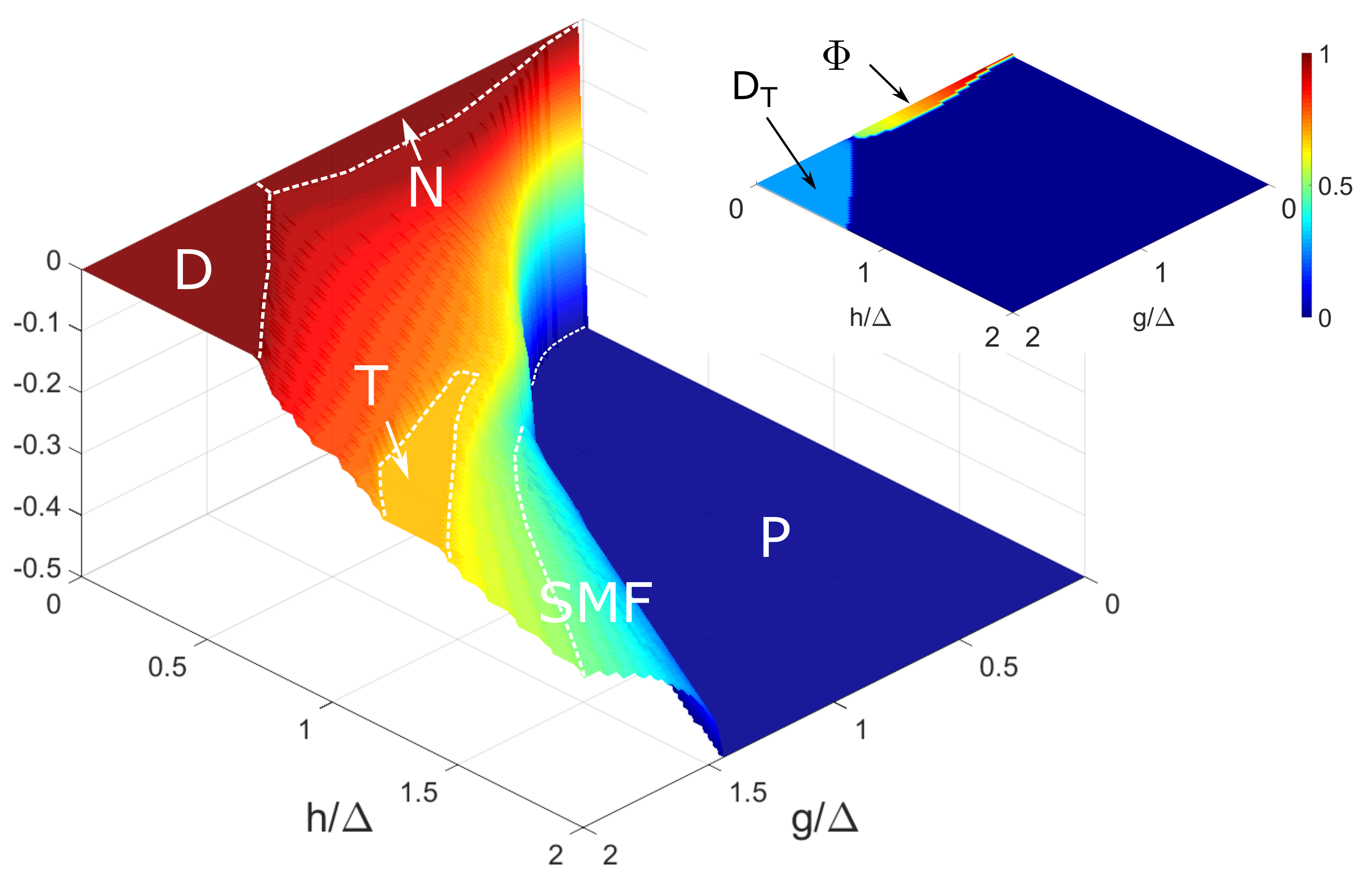}}
	\put(12.5,1){\includegraphics[height=50mm,angle=0,clip]{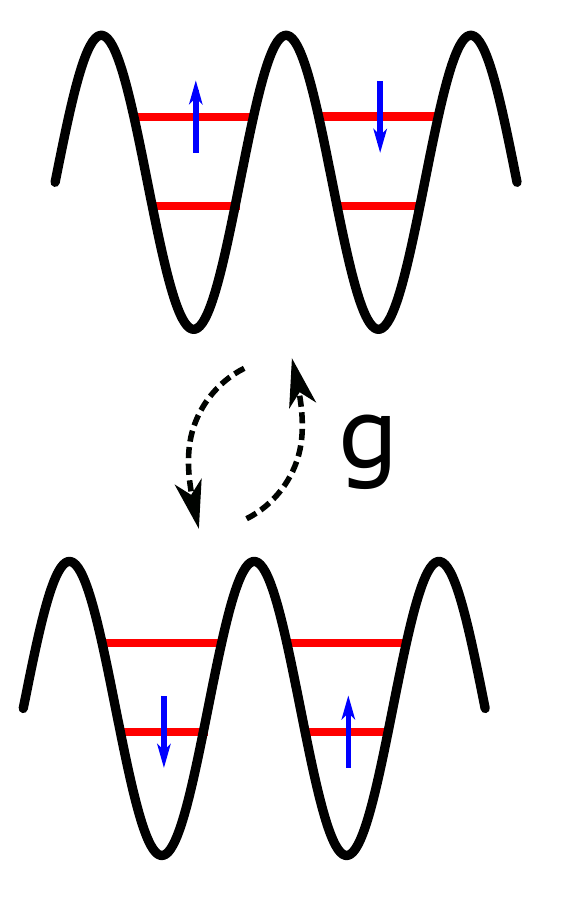}}
	\put(0,15){(a)}
	\put(7,15){(b)}
	\put(0,8){(c)}
	\put(13,8){(d)}
\end{picture}
\caption{\textbf{Quantum spin-motion coupling: model and phase diagram.} (a) Representation of the truncated basis states for few sites. The blue arrow indicates the spin, while the two levels the motional states $\ket{a}$ and $\ket{b}$, separated by an energy difference $\Delta$. (b) Ground state phase diagram obtained studying a system of 62 atoms with open boundary conditions with a DMRG algorithm. We identify unambiguously five phases: a paramagnetic phase (P), a N\'eel ordered phase (N), a dimerized phase of triplets (D), a spin-motion fluid phase (SMF) and a phase of trimers (T). There is an additional phase corresponding to a charge density wave with quasi-long-range order, labeled as SMF(CDW), and whose boundary with a set of still unknown phases U is not well understood. The continuous line is the border of the paramagnetic phase obtained analytically in the weak coupling regime (see text), the dashed line corresponds to $h = -\Delta + 2g$. The 10 red stars indicate parameters $(g,h)$ where the correlations in Fig.~\ref{fig_5}c are evaluated. (c) Surface plot of the magnetization per atom $M_z$, with the phases of (b) indicated. Inset: contour plot of the order parameters $|\Phi|$ and $|D_T|$ in a unique color scale. (d) The virtual process (for $g \ll \Delta$) of two atoms exchanging the spin excitation by jumping to the motional state $\ket{b}$ and returning to the original state, which gives rise to the effective Ising interaction term of Hamiltonian \eqref{Ham_eff}.}
\label{fig_3}
\end{figure}

We now consider a quantum description of motion and spins, which is relevant, e.g., if the motion is initially cooled to its ground state. We proceed by writing the Hamiltonian of Eq.~\eqref{eq:hamiltonian_class_2} in second quantized form, and expanding the spatial field operators in terms of Wannier functions localized at each site (see Appendix B for the details). To make the problem tractable, we will further restrict ourselves to the lowest two motional bands, which we denote by $\ket{a}_i$ and $\ket{b}_i$ as shown in Fig.~\ref{fig_3}a. This represents the minimal model in which spin and motion can couple, since superpositions of states $\ket{a}$ and $\ket{b}$ yield spatial wave-functions that are displaced from the site centers. We also assume that the lattice is deep enough that tunneling is negligible. It will be convenient to introduce a set of pseudo-spin operators on each sit, $\tilde{\sigma}_z= b^\dagger_ib_i - a^\dagger_ia_i$, etc., to represent the motional degree of freedom.

The overall Hamiltonian can thus be expressed in terms of these operators as
\begin{multline}
\label{spin_chain_ham}
H = \sum_i \Delta \tilde{\sigma}^z_i + h \sigma^z_i + 2g\,\bigg\{\tilde{\sigma}^x_i\tilde{\sigma}^x_{i+1} - \frac{1}{2L\eta_0}\bigg[(\eta_a+\eta_b)(\tilde{\sigma}^x_i-\tilde{\sigma}^x_{i+1}) + (\eta_b-\eta_a)(\tilde{\sigma}^x_i\tilde{\sigma}^z_{i+1}-\tilde{\sigma}^z_i\tilde{\sigma}^x_{i+1})\bigg]\bigg\}\times \\
\times\,\big(\sigma^+_i\sigma^-_{i+1} + \text{h.c}\big).
\end{multline}
The terms proportional to $\Delta$ and $h$ describe the energy arising from the band and magnetic field, respectively, while the remainder describes the spin-motion coupling. Here $g = J\,e^{-2a/L}\eta_0^2/2$ is a scaled coupling constant, with $\eta_0 = \int dx\,\sin kx\,w_{a}(x)w_{b}(x)$ and $\eta_{a,b} = \int dx\,x\sin kx\,w^2_{a/b}(x)$ ($w_{a/b}$ being the Wannier functions of states $\ket{a}$ and $\ket{b}$). In the following we take $L = 2a$ and the ratio between the trapping lattice depth $V_L$ and the recoil energy $E_R$ to be 20, for which numerical evaluation of the Wannier functions yields $\eta_0 \approx 0.54$, $\eta_a \approx 0.12a$ and $\eta_b \approx 0.32a$. The terms between brackets contain the dependence of the spin interaction on the motional state of the atoms and have a simple physical explanation. The dominant $\tilde{\sigma}^x_i\tilde{\sigma}^x_{i+1}$ term has largest amplitude when both atoms sit in an equal superposition of states $\ket{a}$ and $\ket{b}$ (i.e., the wave-function is maximally displaced from the center), which reflects that the atoms are trapped at nodes of the APCW. The other terms, which are smaller, originate from the exponentially decaying interaction and are responsible for spatial dimerization. It is interesting to note that this Hamiltonian constitutes an extreme case of spin-orbit coupled systems, as neither an orbital kinetic energy nor a motion-independent spin interaction appear.

We study the phase diagram of Hamiltonian \eqref{spin_chain_ham} in the $g-h$ plane by means of a finite-size density matrix renormalization group (DMRG) algorithm \cite{DMRG}. The resulting phase diagram for $0 \leq g,h \leq 2\Delta$ is shown in Fig. \ref{fig_3}b for $N = 62$ atoms, where we can clearly distinguish at least six phases. The procedure by which these phases are numerically demarcated is discussed in detail in Appendix B, while in the main text we describe the salient physical properties of each phase.
First, for sufficiently large magnetic fields $h > h_\mathrm{crit}(g)$, with $h_\mathrm{crit}(0) = 0$, the spins are fully polarized and thus the spin-motion coupling has no effect. The many-body state is thus separable, with each atom residing in the lowest motional band, $\ket{\psi}=\ket{a,\downarrow}^{\otimes N}$ (``P'' phase in Fig.~\ref{fig_3}b).

Along the $g$-axis up to $g_\mathrm{crit}$ we have a N\'eel ordered phase ``N'', where the magnetization per atom $M_z = 1/(2N)\sum_i \braket{\sigma_i^z}$ is zero and the N\'eel order parameter $\Phi = (1/N)\sum_i(-1)^i\braket{\sigma}^z_i $ has a finite value, as shown in Fig.~\ref{fig_3}c. This phase also extends to finite values of $h$ with a lobe-like shape. The existence of this phase can be predicted analytically in the weak coupling regime, i.e. for $g/\Delta$ small, such that the high-energy excitations associated with populating the upper band can be effectively integrated out. In particular, through a Schrieffer-Wolff transformation \cite{SW} on Eq.~\eqref{spin_chain_ham} one obtains the following effective Hamiltonian acting only on the spin degrees of freedom (see Appendix B):
\begin{equation}
\label{Ham_eff}
H^{\mathrm{eff}} = -N\Delta + \sum_{i}\,h\,\sigma^z_i + J_1\,\big(\sigma^z_i\sigma^z_{i+1} - 1\big) + 2J_2\,\big(\sigma^+_{i-1}\sigma^-_{i+1} + \sigma^-_{i-1}\sigma^+_{i+1}\big).
\end{equation}
Here $J_1 = g^2(1+4\chi^2)/2\Delta$, $J_2 = g^2\chi^2/\Delta$ and $\chi = \eta_a/(\eta_0L)$. Hamiltonian \eqref{Ham_eff} describes a nearest neighbor anti-ferromagnetic (AF) Ising model with an additional XX term coupling next-nearest neighbors, with all such terms mediated by virtual phonons. For example, the spin-motion term in Eq.~\eqref{spin_chain_ham} proportional to $\tilde{\sigma}^x_i \tilde{\sigma}^x_{i+1}$ enables a fluctuation where two consecutive atoms, anti-aligned in their spins, jump to the higher band and exchange their spins, before returning to the original state (see Fig.~\ref{fig_3}d). This process results in a lower energy for the anti-aligned configuration and produces the longitudinal ($\sigma^z_i\sigma^z_{i+1} - 1)$ term in \eqref{Ham_eff}. For zero magnetic field, given that $J_1 \gg J_2$ the ground state exhibits AF ordering along z ($\Phi\approx 1$), as illustrated in Fig.~\ref{fig_3}c. On the other hand, for $h>h_\textrm{crit}(g)$ all spins are in state $\ket{\downarrow}$. Intuitively, one can expect that the transition from N\'eel ordering to polarized occurs with all $\ket{\downarrow}$ spins in the N\'eel phase remaining fixed (subchain ``A''), while the $\ket{\uparrow}$ spins (subchain ``B'') ``melt'' and then re-configure pointing downward. One can thus make an ansatz where subchain A acts as an effective magnetic field for B. Thus, subchain B satisfies an XX model with $H^\textrm{eff}_B = \sum_{i}\,(h-2J_1)\sigma^z_i + 2J_2\,(\sigma^+_{i}\sigma^-_{i+1} + \sigma^-_{i}\sigma^+_{i+1})$, which has the phase diagram represented in Fig.~B.1. With this treatment we can predict $h_\mathrm{crit}(g) \approx g^2/\Delta$ for $g \ll \Delta$ (solid line in Fig.~\ref{fig_3}b), in agreement with the numerics.

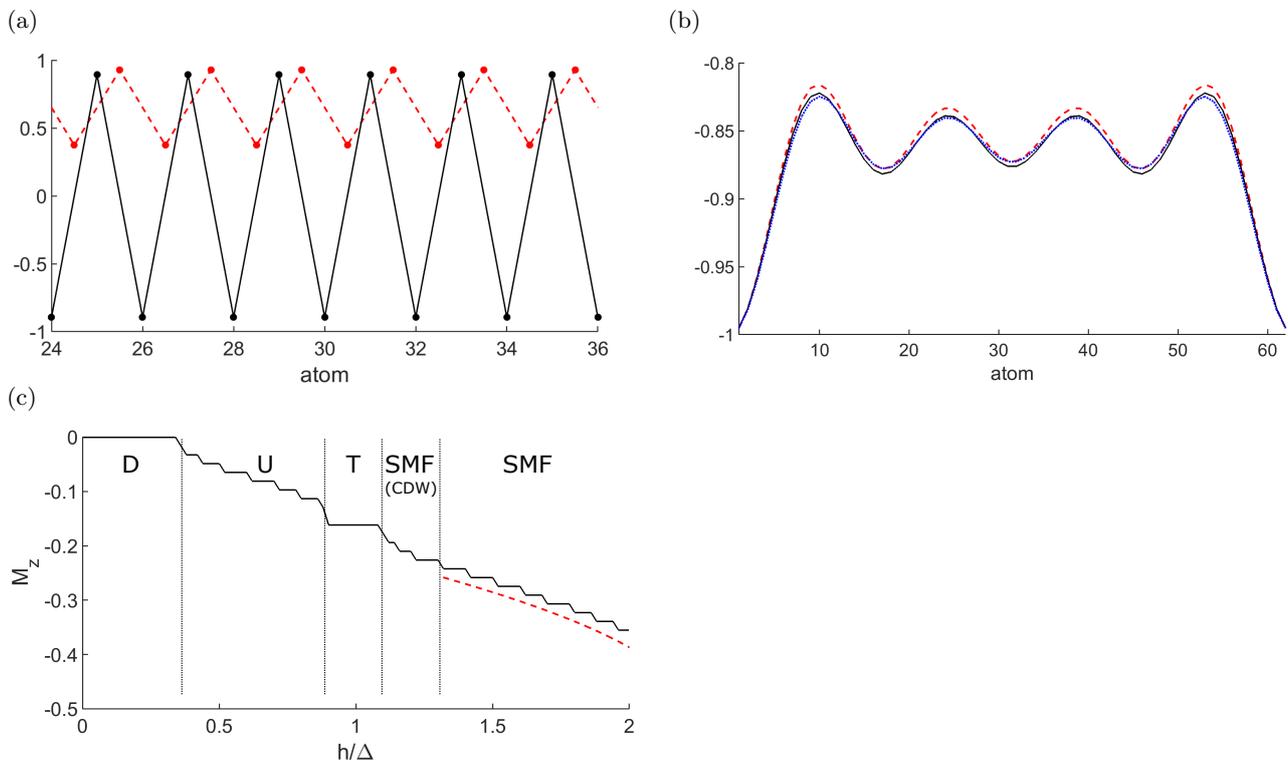
\begin{figure}[t]
	\setlength{\unitlength}{1cm}
	\begin{picture}(20,9)
	\put(0,5){\includegraphics[height=45mm,angle=0,clip]{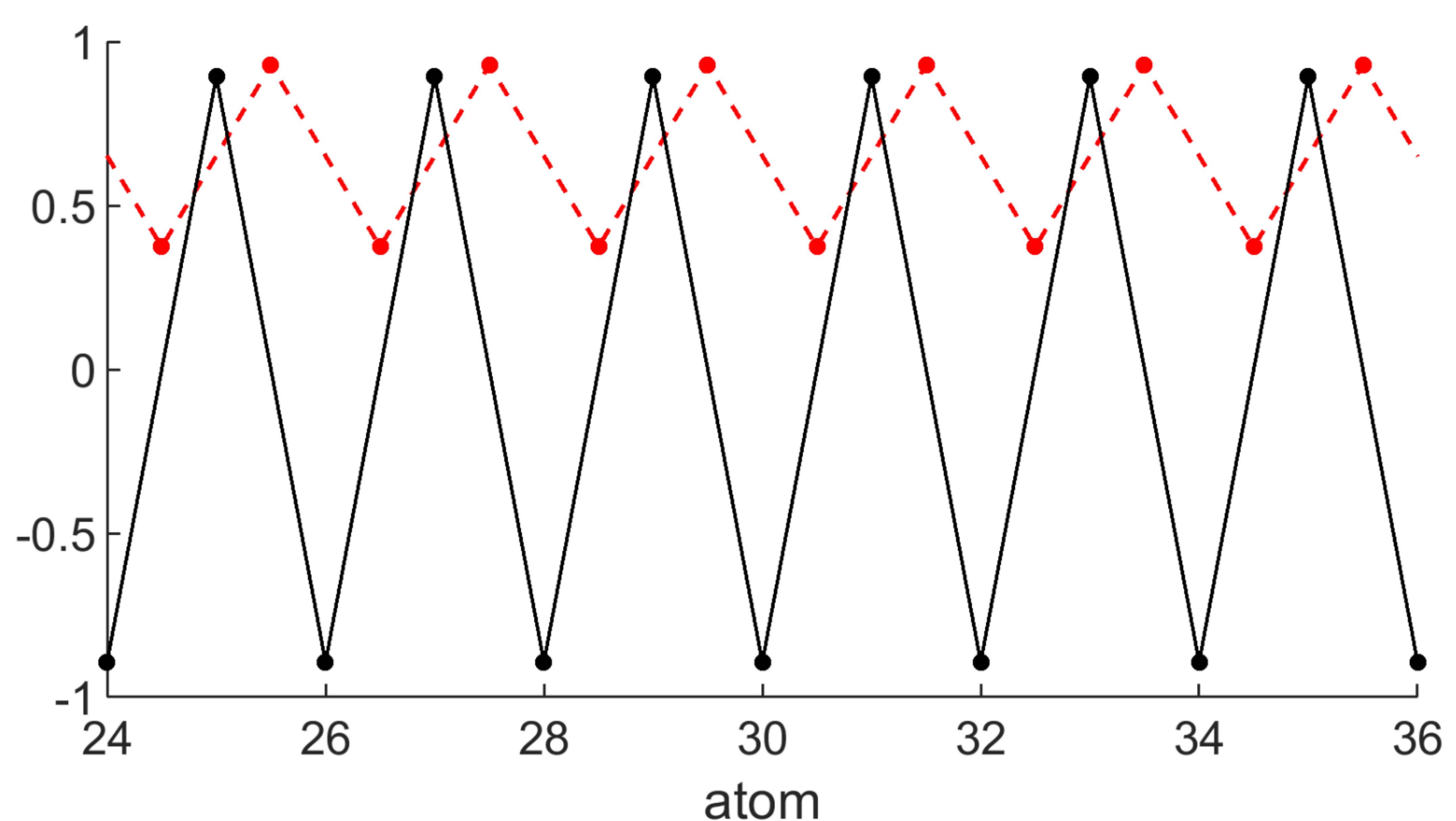}}
	\put(9,5){\includegraphics[height=45mm,angle=0,clip]{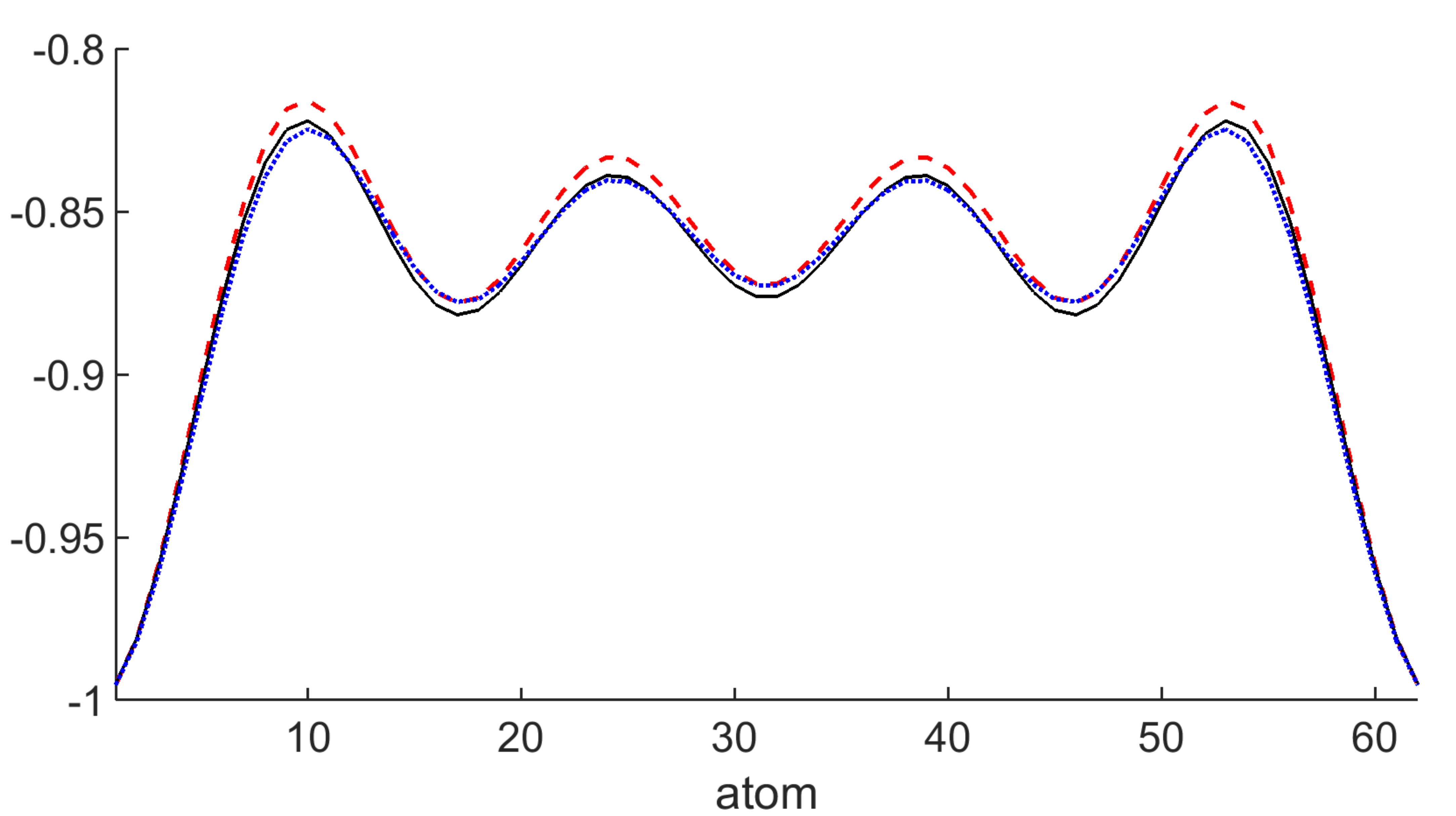}}
	\put(0,0){\includegraphics[height=45mm,angle=0,clip]{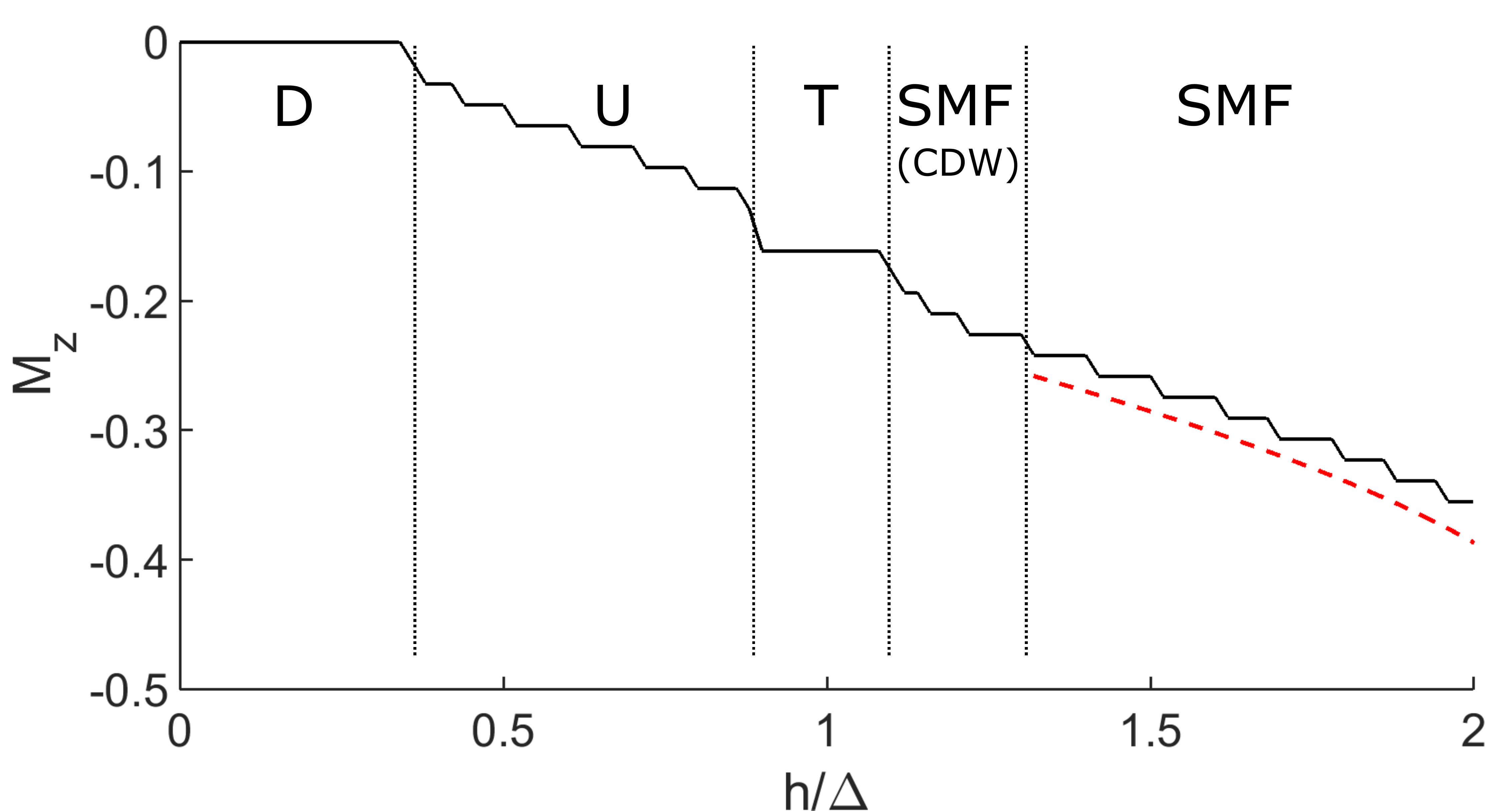}}
	\put(0,9.7){(a)}
	\put(8.7,9.7){(b)}
	\put(0,4.7){(c)}
	\end{picture}
\caption{\textbf{Correlation functions of dimerized and spin-motion fluid phases.} (a) Spin triplet fraction $\braket{T|\rho_{i,i+1}|T}$ (dashed line with dots between $i$ and $i+1$) and displacement $\braket{\tilde{\sigma}^x_i}$ (continuous line) of the ground state for $(g,h) = (1.7,0.2)\Delta$, belonging to the dimerized phase ``D''. Only atoms 24-36 are shown for clarity. (b) $\braket{\sigma^z_i}$ (continuous line), $\braket{\tilde{\sigma}^z_i}$ (dashed line) along the chain for the ground state at $(g,h) = (1.18,1.4)\Delta$ belonging to the spin-motion fluid phase ``SMF". The state contains 4 atoms flipped to $\ket{\uparrow}$ along the direction of the magnetic field. The blue dotted line is $\braket{\tau^z_i}$ on the ground state of $H^+$. (c) Magnetization curve for $g = 1.6\Delta$ as a function of $h$. The dashed line is the magnetization predicted by $H^+$ for the ``SMF" phase.}
\label{fig_4}
\end{figure}

The N\'eel order extends to values of $g/\Delta \gtrsim 1$ where the low-energy description of \eqref{Ham_eff} is no longer accurate, and decreases discontinuously to zero with the onset of a new phase of dimerized triplets (labelled ``D'' in Fig.~\ref{fig_3}b).
This phase is characterized by zero magnetization and a non-zero spin triplet dimer order parameter, defined as $D_T = (1/N)\sum_i (-1)^i \bra{T}\rho_{i,i+1}\ket{T}$ with $\ket{T}$ being the spin triplet state and $\rho_{i,i+1}$ the two-site spin reduced density matrix, as can be seen in Fig. \ref{fig_3}c. It also has a non-zero spatial dimer order parameter, defined as $D_x =  (1/N)\sum_i (-1)^i \braket{\tilde{\sigma}^x_i}$. The entangled dimerized structure is evident in Fig.~\ref{fig_4}a, where we plot the triplet fraction in the two-particle density matrix, $\bra{T}\rho_{i,i+1}\ket{T}$ and the displacement $\braket{\tilde{\sigma}^x_i}$ in a part of the chain for $(g,h) = (1.7,0.2)\Delta$. Also, we can observe that $|\braket{\tilde{\sigma}^x_i}| \sim 1$. Thus, the two-band approximation for the atomic motion is technically violated since the displacement from the trap center is saturated. However, given the classical result, we expect that the prediction of dimerization remains correct.

For simultaneously large values of $g$ and $h$, there is a spin-motion fluid phase (``SMF'') where the system is gapless and the magnetic field strongly polarizes the spins, such that $M_z$ is close to -1/2.
This phase corresponds with good approximation to the ground state of the XX Hamiltonian $H^+ = \sum_i (\Delta+h)\tau^z_i + 2g(\tau^+_i\tau^-_{i+1} + \text{h.c.})$. Here $\tau_i^z$ is the Pauli matrix with eigenstates $\ket{\Downarrow}= \ket{a,\downarrow}$ and $\ket{\Uparrow}=\ket{b,\uparrow}$, while $\tau^{\pm}$ are associated raising and lowering operators. Thus, this phase corresponds to a dilute fluid of \textit{composite flips} of spin and motion. The existence of this phase can be understood by noting that for large magnetic field, the system is only dilutely populated by spins pointing up. Thus the terms in Eq.~\eqref{spin_chain_ham} proportional to $\eta_{a,b}$ that are responsible for dimerization can be neglected.
The structure of the remaining Hamiltonian connects naturally the states $\ket{\Downarrow}$ directly to $\ket{\Uparrow}$, in the form of $H^+$ (see Appendix B for the details). The locking between spin and motional correlations can be observed in Fig.~\ref{fig_4}b, where the expectations values of $\sigma^z_i$ and $\tilde{\sigma}^z_i$ obtained with DMRG are plotted for a representative point in the phase. The oscillations of $\braket{\sigma^z_i}$ and $\braket{\tilde{\sigma}^z_i}$ are due to the open boundary conditions in a finite system and are observable also in a pure XX model. In Fig.~\ref{fig_4}c the magnetization curve predicted by $H^+$ is compared with the numerical result from the DMRG study of the full Hamiltonian for $g = 1.6\Delta$, showing good agreement, while in Fig.~\ref{fig_3}b the predicted boundary with the ``P'' phase $h_\mathrm{crit}(g) \approx -\Delta + 2g$ is represented by a dashed line.

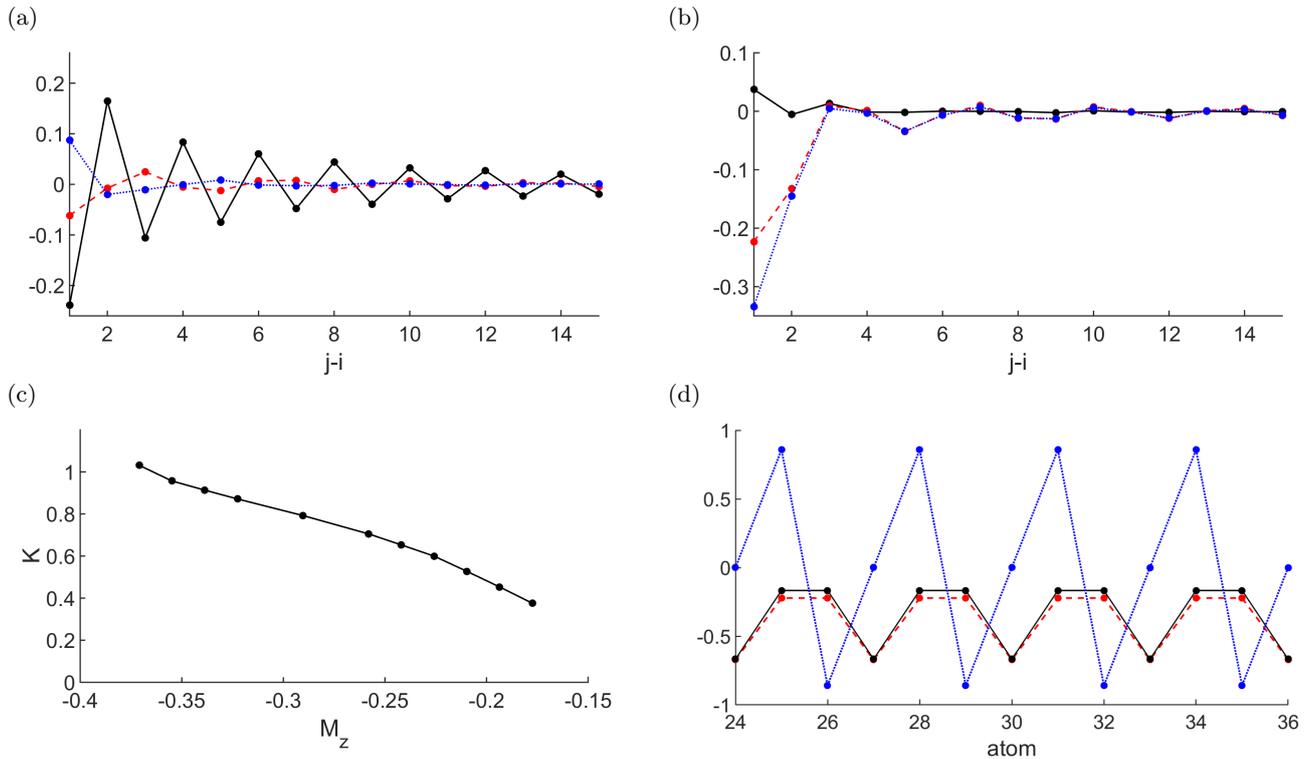
\begin{figure}[t]
	\setlength{\unitlength}{1cm}
	\begin{picture}(20,9.7)
	\put(0,5){\includegraphics[height=45mm,angle=0,clip]{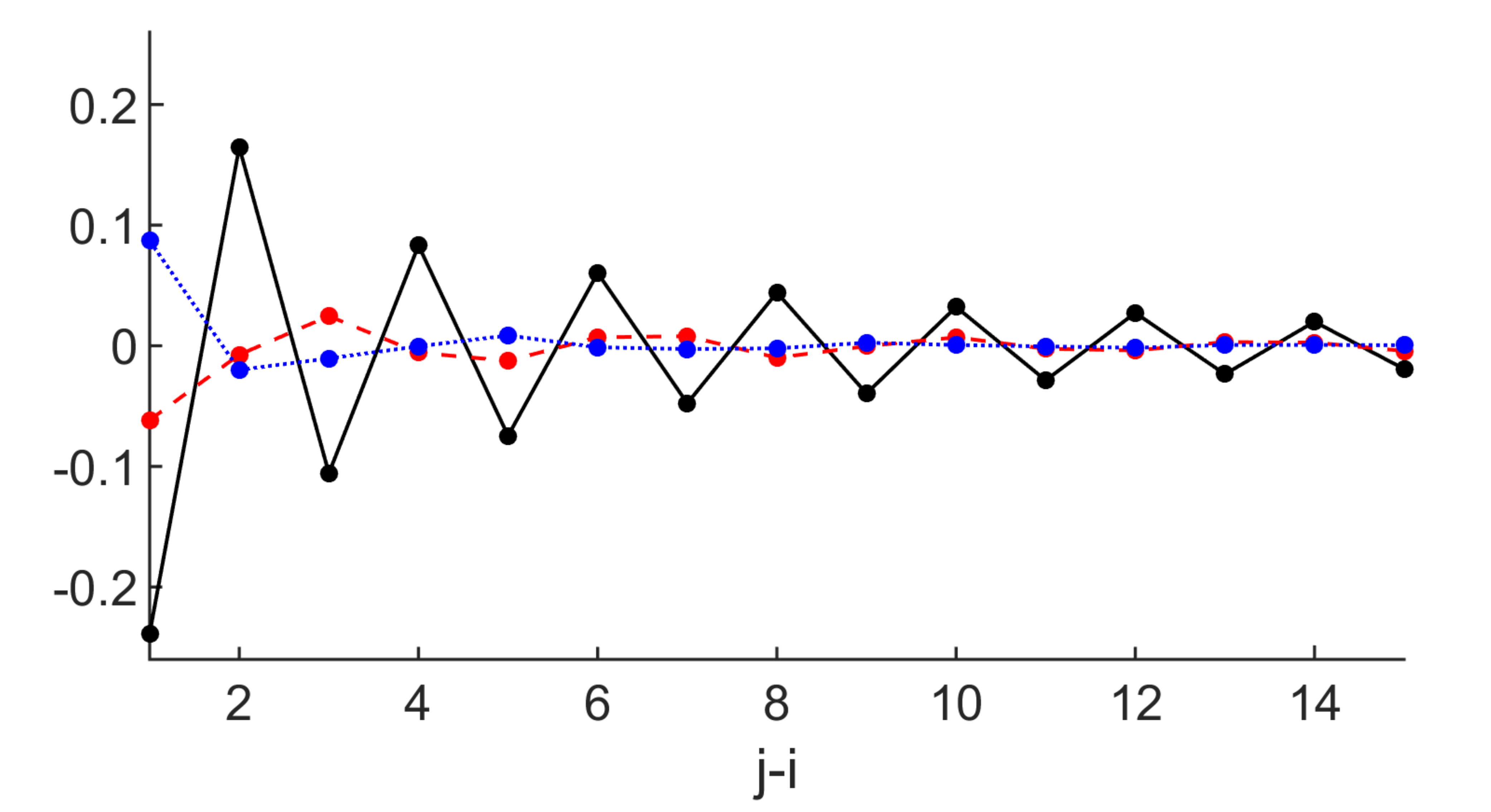}}
	\put(9,5){\includegraphics[height=45mm,angle=0,clip]{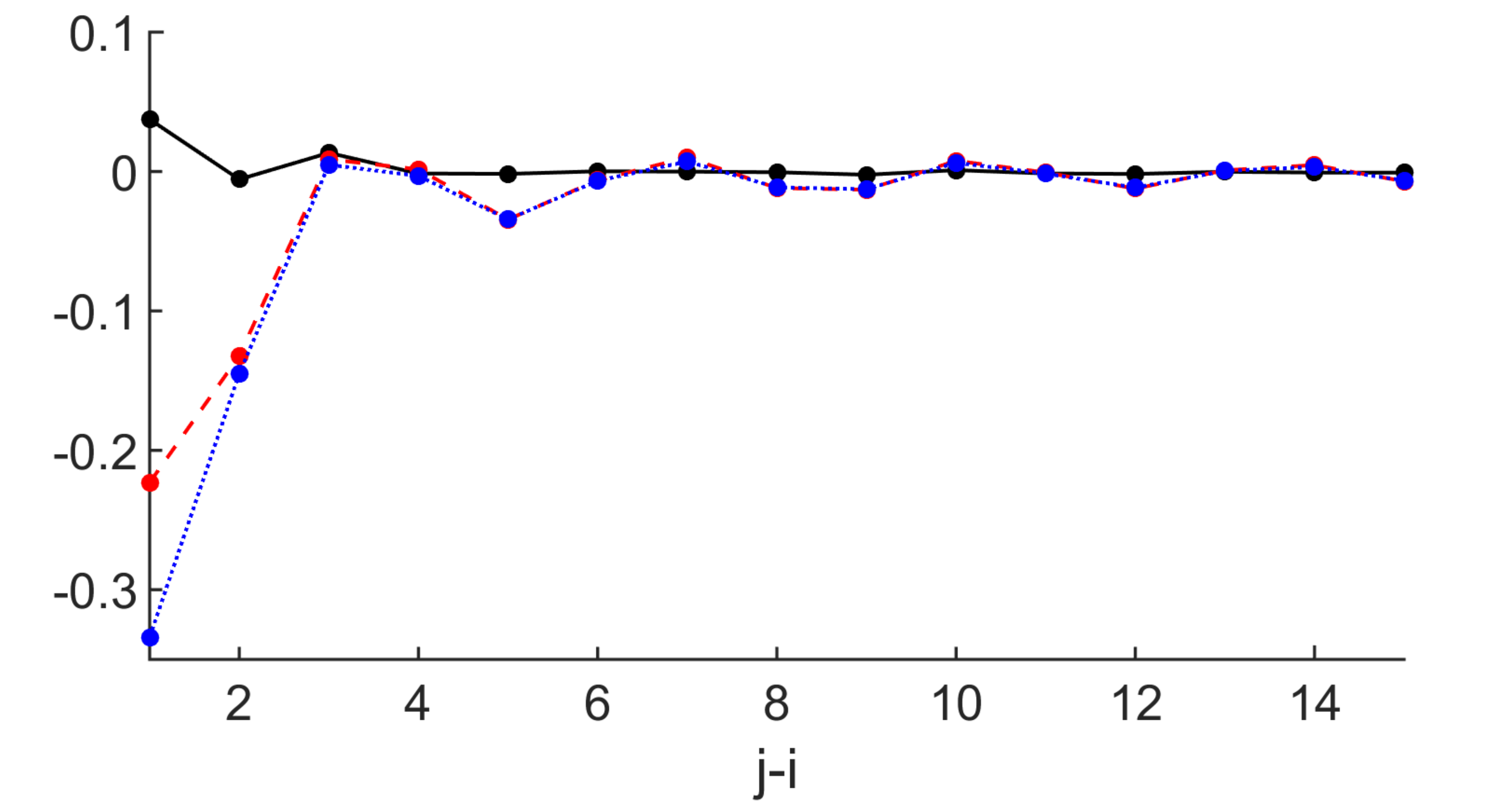}}
	\put(0,0){\includegraphics[height=45mm,angle=0,clip]{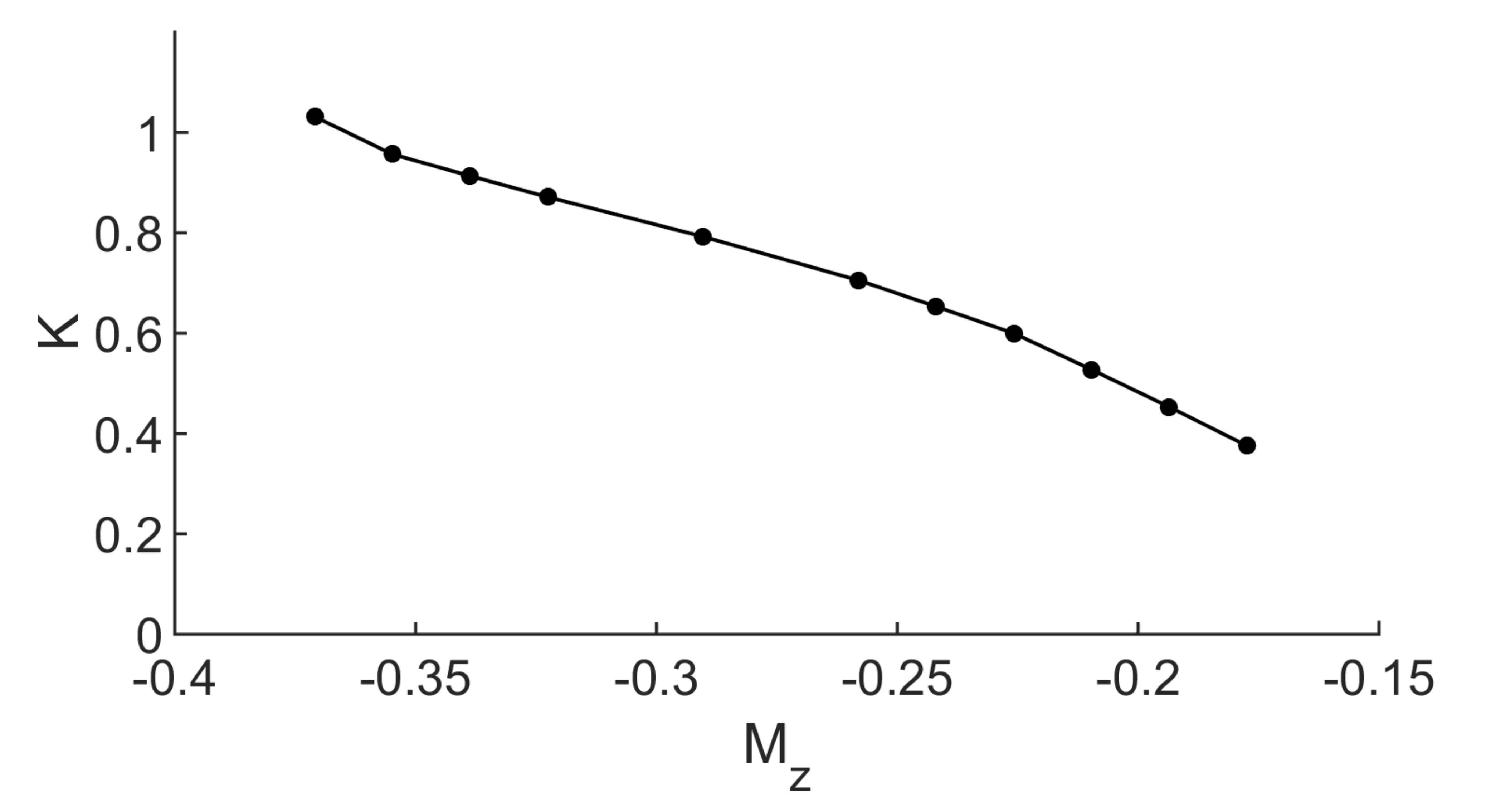}}
	\put(9,0){\includegraphics[height=45.5mm,angle=0,clip]{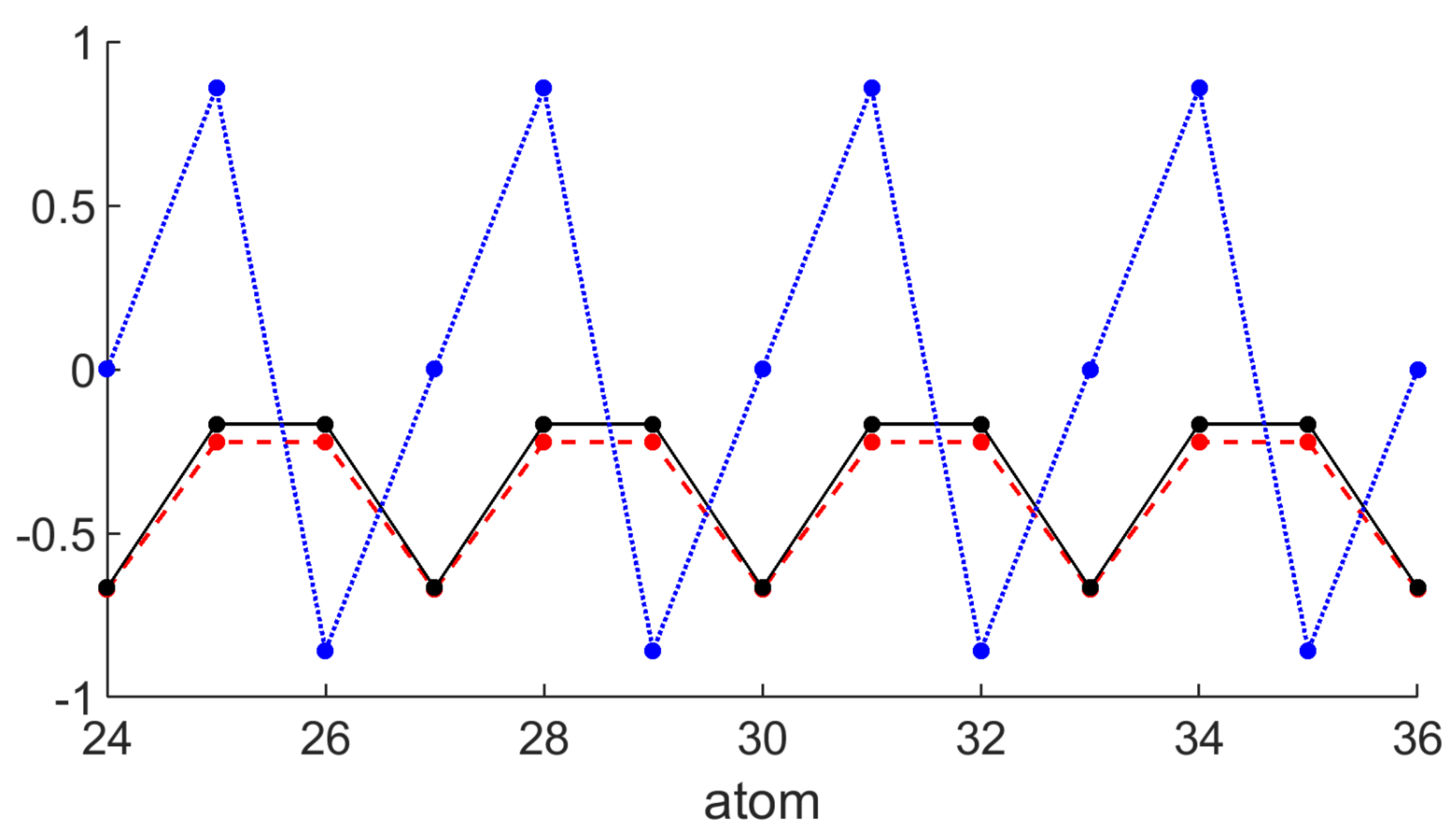}}
	\put(0,9.7){(a)}
	\put(8.7,9.7){(b)}
	\put(0,4.7){(c)}
	\put(8.7,4.7){(d)}
	\end{picture}
\caption{\textbf{Correlation functions of charge density wave and trimer phases.} (a) Correlation functions $\braket{X^+_iX^-_j}-\braket{X^+_i}\braket{X^-_j}$ with $X$ equal to $\tau$ (solid line), $\tilde{\sigma}$ (dashed line) and $\sigma$ (dotted line) at $(g,h) = (1.74,1.38)\Delta$, in the SMF(CDW) region. The value of $i=29$ is taken fixed in the bulk of the chain and $j$ is ranging from 30 to 44. (b) As in (a) but for the density correlation functions $\braket{X^z_iX^z_j}-\braket{X^z_i}\braket{X^z_j}$. (c) Value of $K$ as function of the magnetization, obtained by fitting the numerical results for the $(g,h)$ values marked by stars in Fig.~\ref{fig_3}b. (d) $\braket{\sigma^z_i}$ (continuous line), $\braket{\tilde{\sigma}^z_i}$ (dashed line) and $\braket{\sigma^x_i}$ (dotted line) along the chain for the ground state at $(g,h) = (1.7,1.1)\Delta$, where the ground state belongs to the T phase.}
\label{fig_5}
\end{figure}

For $-1/4 \lesssim M_z < 0$ $H^{+}$ no longer serves as a good description for the ground state. Most of this region consists of a set of phases ``U" whose origin is not completely understood yet. However, for strong interactions $g/\Delta \gtrsim 1$, the system qualitatively appears to behave as an interacting Luttinger liquid for the $\tau$ particles. Numerical evidence is shown in Fig.~\ref{fig_5}a, where the two-point correlation functions $C^X_{ij} \equiv \braket{X^+_iX^-_j}-\braket{X^+_i}\braket{X^-_j}$ are plotted for various $X=\tau,\sigma,\tilde{\sigma}$. In particular, $C^\tau_{ij} \sim (-1)^{|j-i|}|j-i|^{-1/2K}$, exhibits a power law decay, while correlations of the other degrees of freedom exhibit more erratic behavior. Similar observations hold for the density correlation functions (Fig.~\ref{fig_5}b). We fit the Luttinger parameter $K$ \cite{giamarchi} from the numerical data for $C_{ij}^{\tau}$, taking the points in Fig.~\ref{fig_3}b across the SMF to U boundary, and plot these points in Fig.~\ref{fig_5}c. The decrease below $K=1$ is indicative of the formation of a charge density wave phase with quasi-long-range order, i.e. algebraic decay of the correlation functions. We thus denote this part of the phase diagram as SMF(CDW). The determination of the precise boundary of this phase will be investigated in future work.

The decrease in $K$ is also known to facilitate the possibility of phases with spontaneously broken symmetry, which is observed in our system as well. In particular, for $M_z \rightarrow -1/6$, $K$ tends to zero, i.e. the decay of $C^\tau_{x}$ turns into exponential. At $M_z = -1/6$ (one third of the maximum magnetization), we observe the presence of a plateau in the magnetization curve (Figs.~\ref{fig_3}c and \ref{fig_4}c), for values of $g$ sufficiently large. In this region the ground state assumes a trimerized configuration, as shown in Fig.~\ref{fig_4}d, where $\braket{\sigma^z_i}, \braket{\tilde{\sigma}^z_i}$ and the displacement $\braket{\tilde{\sigma}^x_i}$ are plotted. 
While we are not able to predict the appearance of such a plateau in our model from first principles, we note that all of its features are consistent with the conditions of Ref.~\cite{affleck}. In particular, our Hamiltonian allows for a gapped phase with spontaneously broken symmetry in the ground state with spatial periodicity $n=3$, provided that the quantization condition $n(S-M_z)= $ integer is satisfied (here $S=1/2$ is the total spin). Such a gapped phase should be accompanied by a magnetization plateau.

\section{Conclusions and outlook}

The platform of cold atoms coupled to photonic crystals offers fascinating opportunities to create quantum materials in which spin and motion interact strongly with one another. We have analysed in detail the ground state properties of one experimentally feasible setup, but there exist many exciting avenues for future research. For example, we expect that such systems should exhibit an interesting excitation spectrum, including the possibility of excitations carrying fractional spin \citep{tonegawa}. It would also be interesting to understand the transport properties when spin and motion strongly hybridize. Moreover, it would be highly interesting to consider models without an external lattice potential, and investigate whether the spin interaction alone can produce full spin-entangled crystallization. One might also consider models where the spin part of the interaction already exhibits non-trivial character, such as frustration or topology. Finally, in terms of applications, it would be interesting to explore whether specially engineered spin-motion Hamiltonians can give rise to useful many-body spin states (such as squeezed states for metrology), when the spin interaction alone is incapable of producing such states.

\textit{Acknowledgements} The authors acknowledge A.~M. Rey, J. Alicea, G. Refael, and H.~J. Kimble for stimulating discussions, and the Centro de Ciencias de Benasque ``Pedro Pascual", where this work was initiated. MTM thanks the Hamburg Institut f\"ur Laserphysik, for their hospitality in an extended research visit. The authors thank Jutho Heageman for writing the DMRG code used and making it available to them through the Ghent Tensor Network Summer School. This work was supported by ERC Starting Grant FOQAL, the MINECO Plan Nacional Grant CANS, the MINECO Severo Ochoa Grant SEV-2015-0522 and ``la Caixa-Severo Ochoa'' PhD Fellowship.

\newpage
\appendix
\renewcommand\thefigure{\thesection.\arabic{figure}}  
\setcounter{figure}{0}    

\section{Classical motion}

Here, we provide additional details of the calculations related to Section III of the main text.

\subsection*{Hamiltonian diagonalization}

Hamiltonian \eqref{hamiltonian_class} in the nearest-neighbor approximation and under the ansatz
\begin{equation}
\delta_i = (-1)^i\delta
\end{equation}
on the atomic positions, becomes
\begin{equation}
\label{hamiltonian_class_nn}
H = \frac{V_L}{2}N\,\sin^2 k\delta/2 - J\sum_{i}\,\sin^2 k\delta\,\exp[-2(a+(-1)^i\delta)/L]\,(\sigma^+_i\sigma^-_{i+1} + \text{h.c}) + h\,\sigma^z_i.
\end{equation}
The spin operators can be mapped to fermion annihilation and creation operators $c_i, c_i^{\dagger}$ using the using Jordan-Wigner transformation \cite{JW}
\begin{eqnarray}
\label{J-W}
\sigma_i^+ &=&c^\dagger_i\,e^{\ii\pi\sum_{l<i}c^\dagger_lc_l}, \\
\sigma_i^-&=&e^{-\ii\pi\sum_{l<i}c^\dagger_lc_l}\,c_i, \\
\sigma^z_i&=&2c^\dagger_ic_i - 1,
\end{eqnarray}
which transforms \eqref{hamiltonian_class_nn} into 
\begin{equation}
\label{hamiltonian_class_nn_ferm}
H = E^\mathrm{tr}(\delta) - J\sum_{i}\,\sin^2 k\delta\,\exp[-2(a+(-1)^i\delta)/L]\,(c^\dagger_ic_{i+1} + \text{h.c}) + h\,(2c^\dagger_ic_i-1),
\end{equation}
where $E^\mathrm{tr}(\delta) = NV_L/2\,\sin^2 k\delta/2$.
We note that, because of the dimerization ansatz, two different couplings appear now in the Hamiltonian:
\begin{eqnarray}
\label{coupling}
J_S(\delta)&=&J\,\sin^2 k\delta\,\exp[-2(a-\delta)/L], \\
J_W(\delta)&=&J\,\sin^2 k\delta\,\exp[-2(a+\delta)/L].
\end{eqnarray}
Furthermore, it is natural to relabel the atoms by dimerized pairs (indexed by ``$j$") and the position of the atom inside the pair, i.e. left or right. Then Hamiltonian \eqref{hamiltonian_class_nn} can be expressed as 
\begin{equation}
\label{hamiltonian_class_nn_ferm_2}
H = E^\mathrm{tr}(\delta) - J_S(\delta)\,\sum_{j}\,(c^\dagger_{j,L}c_{j,R} + \text{h.c}) - J_W(\delta)\,\sum_{j}\,(c^\dagger_{j,R}c_{j+1,L} + \text{h.c}) +  2h\,\sum_j\,(c^\dagger_{j,L}c_{j,L}+c^\dagger_{j,R}c_{j,R}-1).
\end{equation}
In this way we have divided the interactions between atoms in the same dimer and atoms in neighboring dimers.

We perform now a change of basis in each dimer, introducing the fermionic operators
\begin{eqnarray}
&s_j&= \frac{1}{\sqrt{2}}(c_{L,j} - c_{R,j}), \\
&t_j&= \frac{1}{\sqrt{2}}(c_{L,j} + c_{R,j}) 
\end{eqnarray}
in terms of which \eqref{hamiltonian_class_nn_ferm_2} is 
\begin{multline}
\label{hamiltonian_class_nn_ferm_3}
H = E^\mathrm{tr}(\delta) - J_S(\delta)\,\sum_{j}\,(t^\dagger_{j}t_{j} - s^\dagger_{j}s_{j}) - \frac{J_W(\delta)}{2}\,\sum_{j}\,(t^\dagger_{j}t_{j+1} - s^\dagger_{j}s_{j+1} + t^\dagger_{j}s_{j+1} - s^\dagger_{j}t_{j+1} + \text{h.c}) + \\ +  2h\,\sum_j\,(t^\dagger_{j}t_{j}+s^\dagger_{j}s_{j}-1). 
\end{multline}
In Fourier space \eqref{hamiltonian_class_nn_ferm_3} becomes 
\begin{equation}
\label{hamiltonian_class_nn_ferm_4}
H =  E^\mathrm{tr}(\delta) - Nh - \sum_{q} 
\begin{pmatrix}
t_q^\dagger & s_q^\dagger
\end{pmatrix}
\begin{pmatrix}
J_S(\delta) + J_W(\delta)\,\cos q - 2h& +\ii J_W(\delta)\,\sin q \\
-\ii J_W(\delta)\,\sin q & -J_S(\delta) - J_W(\delta)\,\cos q - 2h \\
\end{pmatrix}
\begin{pmatrix}
t_q \\
 s_q
\end{pmatrix},
\end{equation}
which can be easily diagonalized as 
\begin{equation}
\label{hamiltonian_class_diag}
H = E^\mathrm{tr}(\delta) - Nh + \sum_{q}\,\bigg[(2h+\epsilon_q)\,d^\dagger_qd_q + (2h-\epsilon_q)\,u^\dagger_qu_q\bigg]. 
\end{equation}
The spectrum is given by 
\begin{equation}
\epsilon_q(\delta)= \bigg(J^2_S(\delta) + J^2_W(\delta) + 2J_S(\delta)J_W(\delta)\cos q\bigg)^{1/2} = J\,e^{-2a/L}\,\sin^2 k\delta\,\bigg(4\cosh^2 2\delta/L + 2(\cos q-1)\bigg)^{1/2}
\end{equation}
and
\begin{eqnarray}
\label{ud}
&d_q&= \frac{1}{\sqrt{(\epsilon_q+a_q)^2+b_q^2}}\bigg(\ii(\epsilon_q+a_q)t_{q} + b_q s_{q}\bigg), \\ 
&u_q&= \frac{1}{\sqrt{(\epsilon_q-a_q)^2+b_q^2}}\bigg(-\ii(\epsilon_q-a_q)t_{q} +  b_q s_{q}\bigg), 
\end{eqnarray}
with $a_q = J_S(\delta) + J_W(\delta)\,\cos q$ and $b_q =  J_W(\delta)\,\sin q$. 

Since $\epsilon_q$ is positive for every $q$ and $J$ has been assumed to be positive, the ground state involves only $u$ operators and is equal to 
\begin{equation}
\label{gs}
\ket{GS}_\delta = \bigg(\prod_{q|\epsilon_q(\delta)>2h}u^\dagger_q\bigg)\ket{0}.
\end{equation}
The ground state energy per atom is:
\begin{eqnarray}
\label{nn_gse}
E(\delta)&=& E^\mathrm{tr}(\delta) - h -\frac{1}{N}\sum_{q|\epsilon_q(\delta)>2h}\,2h-\epsilon_q(\delta). 
\end{eqnarray}
The interaction energy per atom (i.e., the term proportional to $J$) has an analytical expression for $h = 0$, given by
\begin{equation}
E^\mathrm{int}_{h=0}(\delta) = -\frac{2J\,e^{-2a/L}}{\pi}\,\sin^2 k\delta\,\cosh 2\delta/L\,E(\cosh^{-1} 2\delta/L).
\end{equation}
Here we have taken the thermodynamic limit and replaced the summation on $q$ by an integral and $E(k)$ denotes the complete elliptic integral of the second kind.

\subsection*{Triplet fraction}

From \eqref{gs} we can calculate the triplet (and singlet) fraction for two atoms within a dimer, which is defined as
\begin{equation}
\label{triplet_frac}
T_S(\delta) = \frac{1}{N}\sum_j \braket{t^\dagger_jt_j(1-s^\dagger_js_j)} = \frac{1}{N}\sum_j \braket{t^\dagger_jt_j} -\braket{s^\dagger_j s_jt^\dagger_jt_j},
\end{equation}
where the expectation value is taken on the ground state. For $h=0$ we can calculate it analytically. Inverting \eqref{ud} and using the expression for $t$ and $s$ as function of $u$ and $d$ we find that 
\begin{equation}
T_S^{h=0}(\delta) = \bigg(\frac{1}{2} + I_S(\delta)\bigg)^2,
\end{equation}
where 
\begin{equation}
I_S(\delta) = \frac{1}{2\pi}\int^{\pi}_{-\pi} dq\,\frac{a_q}{2\epsilon_q}. 
\end{equation}
$I_S(\delta)$ tends to $1/\pi$ for $\delta \rightarrow 0$.
Similarly we can calculate the triplet fraction for consecutive atoms in different dimers
\begin{equation}
\label{triplet_frac}
T_W(\delta) = \frac{1}{N}\sum_j \braket{t'^\dagger_jt'_j(1-s'^\dagger_js'_j)} = \frac{1}{N}\sum_j \braket{t'^\dagger_jt'_j} -\braket{s'^\dagger_j s'_jt'^\dagger_jt'_j},
\end{equation}
with $t'_j = 1/2(t_j-s_j+t_{j+1}+s_{j+1})$ and $s'_j = 1/2(t_j-s_j-t_{j+1}-s_{j+1})$, which for $h = 0$ is equal to  
\begin{equation}
T_W^{h=0}(\delta) = \bigg(\frac{1}{2} + I_W(\delta)\bigg)^2,
\end{equation}
where 
\begin{equation}
I_W(\delta) = \frac{1}{2\pi}\int^{\pi}_{-\pi} dq\,\frac{a_q\cos q+b_q\sin q}{2\epsilon_q}. 
\end{equation}
$I_W(\delta)$ tends to $1/\pi$ for $\delta \rightarrow 0$.

\section{Quantum motion}

Here we provide supplemental calculations and discussion to Section IV in the main text.

\subsection*{Hamiltonian in the two-band approximation}

In this sub-section, we derive the Hamiltonian given by Eq.~\eqref{spin_chain_ham} in the main text, describing spin-motion coupling in the two-band approximation. We begin by writing the full Hamiltonian of Eq. \eqref{hamiltonian_class} in the main text in second quantization, introducing the annihilation operator  $\Psi_{\sigma}(x)$, where $\sigma$ denotes the spin state. The Hamiltonian then reads
\begin{equation}
\label{eq:hamiltonian_quantum}
H = \sum_\sigma\,\int dx\,\Psi_{\sigma}^\dagger(x)\big[-\frac{\hbar^2}{2m}\nabla^2 + V^\mathrm{trap}(x)\big]\Psi_{\sigma}(x) + \sum_{\sigma,\sigma'}\,\int dxdx'\,\Psi_{\sigma}^\dagger(x)\Psi_{\sigma'}^\dagger(x')V^\mathrm{int}_{\sigma,\sigma'}(x,x')\Psi_{\sigma'}(x')\Psi_{\sigma}(x),
\end{equation}
where $ V^\mathrm{int}_{\sigma,\sigma'}(x,x') = J\,e^{-|x-x'|/L}\,E(x)E(x')\,(\sigma^{+}_{\sigma}\sigma^{-}_{\sigma'} + \text{h.c})$ is the spin-dependent interaction discussed in Section II of the main text.

The field operators can be expanded in the basis of the Wannier functions (which we assume to be real without loss of generality) as $\Psi_\sigma(x) = \sum_i\sum_{\alpha}\,w_{i,\alpha}(x)\psi_{i,\alpha,\sigma}$, where $i$ and $\alpha$ are the site and band index, respectively. Here $w(x)$ denotes the spatial wavefunction associated with the Wannier function, and $\psi$ its associated annihilation operator. As a minimal model to investigate the physics of spin-motion coupling, we will truncate the system to the two lowest bands, i.e. $\Psi_\sigma(x) \approx \sum_{i}\,w_{i,a}(x)\,a_{i,\sigma} + w_{i,b}(x)\,b_{i,\sigma}$, where $a_i$ and $b_i$ are the annihilation operators for the lowest and first excited bands, respectively. The non-interacting Hamiltonian in this basis is 
\begin{multline}
H_\mathrm{0} = \sum_{i,j}\sum_{\sigma}\int dx\,\big[w_{a,i}(x)a_{i,\sigma}^\dagger + w_{b,i}(x)b_{i,\sigma}^\dagger\big] \bigg[-\frac{\hbar^2}{2m}\nabla^2 + V^\mathrm{tr}(x)\bigg]\big[w_{a,j}(x)a_{j,\sigma} + w_{b,j}(x)b_{j,\sigma}\big] \approx \\ \approx \sum_{i}\sum_{\sigma}\bigg(-t_a a^\dagger_{i,\sigma}a_{i+1,\sigma} - t_b b^\dagger_{i,\sigma}b_{i+1,\sigma} + \text{h.c.}\bigg) + \Delta b^\dagger_{i,\sigma}b_{i,\sigma}.
\end{multline}
The tunneling coefficients are defined in the usual way as
\begin{equation}
t_\alpha = -\int dx\,w_{\alpha,i}(x)\bigg[-\frac{\hbar^2}{2m}\nabla^2 + V^\mathrm{tr}(x)\bigg]w_{\alpha,i+1}(x).
\end{equation}

The long-range spin-dependent interaction term has a kind of cumbersome expression in second quantization. The first simplification in the brute force expansion is to neglect terms with coefficients containing expressions of the form $w_{i,\alpha}(x)w_{j,\alpha'}(x)$ with $i\neq j$, since the overlap of Wannier functions on different sites is small. After this simplification we have 
\begin{multline}
\label{hamil_exp}
H^{\mathrm{int}} = \frac{J}{2}\,\sum_{i,j}\bigg\{V^{ij}_aa^\dagger_{i,\uparrow}a^\dagger_{j,\downarrow}a_{j,\uparrow}a_{i,\downarrow} + V^{ij}_bb^\dagger_{i,\uparrow}b^\dagger_{j,\downarrow}b_{j,\uparrow}b_{i,\downarrow} + \\ + V^{ij}_{ab}\big(a^\dagger_{i,\uparrow}b^\dagger_{j,\downarrow}b_{j,\uparrow}a_{i,\downarrow} + b^\dagger_{i,\uparrow}a^\dagger_{j,\downarrow}a_{j,\uparrow}b_{i,\downarrow}\big) + {V^{ij}_{ab}}'\big(a^\dagger_{i,\uparrow}a^\dagger_{j,\downarrow}b_{j,\uparrow}b_{i,\downarrow} + b^\dagger_{i,\uparrow}b^\dagger_{j,\downarrow}a_{j,\uparrow}a_{i,\downarrow} + \\
+ a^\dagger_{i,\uparrow}b^\dagger_{j,\downarrow}a_{j,\uparrow}b_{i,\downarrow} + b^\dagger_{i,\uparrow}a^\dagger_{j,\downarrow}b_{j,\uparrow}a_{i,\downarrow}\big) +  V^{ij}_{3ab}\big(a^\dagger_{i,\uparrow}a^\dagger_{j,\downarrow}a_{j,\uparrow}b_{i,\downarrow} + a^\dagger_{i,\uparrow}a^\dagger_{j,\downarrow}b_{j,\uparrow}a_{i,\downarrow} + \\
+ a^\dagger_{i,\uparrow}b^\dagger_{j,\downarrow}a_{i,\uparrow}a_{i,\downarrow} + b^\dagger_{i,\uparrow}a^\dagger_{j,\downarrow}a_{i,\uparrow}a_{i,\downarrow}\big) +  V^{ij}_{3ba}\big(b^\dagger_{i,\uparrow}b^\dagger_{j,\downarrow}b_{j,\uparrow}a_{i,\downarrow} + b^\dagger_{i,\uparrow}b^\dagger_{j,\downarrow}a_{j,\uparrow}b_{i,\downarrow} + \\
+ b^\dagger_{i,\uparrow}a^\dagger_{j,\downarrow}b_{j,\uparrow}b_{i,\downarrow} + a^\dagger_{i,\uparrow}b^\dagger_{j,\downarrow}b_{j,\uparrow}b_{i,\downarrow}\big) + \text{h.c.}\bigg\}.
\end{multline} 
Using the fact that $w_{i,\alpha}(x) = w_{0,\alpha}(x-x_i)$ and that the center of the trap sites $x_i$ coincide with even nodes of the photonic crystal Bloch function, the coefficients $V$ take the form
\begin{eqnarray}
V^{ij}_{\alpha}&=&\int dxdx'\,\sin kx\sin kx'\,e^{-|x_i-x_j-x+x'|/L}\,w^2_{\alpha}(x)w^2_{\alpha}(x'), \\
V^{ij}_{ab}&=&\int dxdx'\,\sin kx\sin kx'\,e^{-|x_i-x_j-x+x'|/L}\,w^2_{a}(x)w^2_{b}(x'), \\
{V^{ij}_{ab}}'&=&\int dxdx'\,\sin kx\sin kx'\,e^{-|x_i-x_j-x+x'|/L}\,w_{a}(x)w_{b}(x)w_{a}(x')w_{b}(x'), \\
V^{ij}_{3\alpha\alpha'}&=&\int dxdx'\,\sin kx\sin kx'\,e^{-|x_i-x_j-x+x'|/L}\,w^2_{\alpha}(x)w_{\alpha}(x')w_{\alpha'}(x'),
\end{eqnarray}
where the site dependence only appears in the exponential.

These coefficients can be greatly simplified by assuming that the Wannier functions are tightly confined around the lattice sites, such that $|x-x'|<L$ in the region over which the Wannier functions have appreciable weight. This motivates an expansion of the coefficients in powers of $1/L$. Furthermore, we will assume that $L \sim a$, such that we can make the nearest-neighbor approximation. Further exploiting the parity of the functions $w_a$ and $w_b$ and the sine function, one finally arrives at
\begin{equation}
\label{H_0}
H^{\mathrm{int}\,(0)} = 2g\,\sum_i\,\tilde{\sigma}^x_i\tilde{\sigma}^x_{i+1}\,\big(\sigma^+_i\sigma^-_{i+1} + \text{h.c.}\big),
\end{equation}
\begin{equation}
\label{H_1}
H^{\mathrm{int}\,(1)} = -\frac{2g}{2L\eta_0}\,\sum_i\,\bigg\{(\eta_a+\eta_b)\big(\tilde{\sigma}^x_i - \tilde{\sigma}^x_{i+1}\big) + (\eta_b-\eta_a)\big(\tilde{\sigma}^x_i\tilde{\sigma}^z_{i+1} - \tilde{\sigma}^x_{i+1}\tilde{\sigma}^z_{i+1}\big)\bigg\}\,\big(\sigma^+_i\sigma^-_{i+1} + \text{h.c.}\big), 
\end{equation}
where $H^{\mathrm{int}\,(n)}$ denote the expansions of the Hamiltonian in powers of $L^{-n}$. Here we have assumed unitary occupation of the trapping sites and introduced the operators with the tilde for the band state as discussed in the main text and $g = Je^{-2a/L}\,\eta_0^2/2$, $\eta_0 = \int dx\,\sin k_0x\,w_{a}(x)w_{b}(x)$ and $\eta_{a,b} = \int dx\,x\sin k_0x\,w^2_{a/b}(x)$. Hamiltonian \eqref{spin_chain_ham} of the main text, in which the tunneling terms are neglected, is given by the sum of \eqref{H_0}, \eqref{H_1}, $H^\mathrm{band} = \Delta\sum_i \tilde{\sigma}^z_i$ and $H^\mathrm{magn} = h\sum_i\sigma_i^z$.

\subsection*{DMRG algorithm}

The DMRG algorithm is a well-established numerical method for ground state studies of one-dimensional systems \cite{DMRG}. It consists of approximating the exact ground state $\ket{GS} = \sum_{i_1,..,i_N} c_{i_1,..,i_N}\ket{i_1,..,i_N}$ of a finite-size system of $N$ sites with local dimension $d$ with a matrix product state (MPS), i.e. a state of the form
\begin{equation}
\label{MPS}
\ket{GS}_{MPS} = \sum_{i_1,..,i_N}\,A^{[1]}_{i_1}A^{[2]}_{i_2}..A^{[N]}_{i_N}\ket{i_1,..,i_N},
\end{equation}
where the matrices $A$ have maximum dimension $D$. The DMRG algorithm finds the matrices $\{A^{[k]}_{i_k}\}$ for which $\braket{H}$ is minimum. This is obtained in an iterative way by optimizing a single matrix in every step, keeping constant all the others and shifting one site at every step. In $2N$ steps the system has been ``swept" once. The level of approximation can be determined by the relative energy difference d$E$ after a sweep. In general for a gapped Hamiltonian, with $D = 100-1000$ the energy converges in few sweeps and one obtains an extremely good approximation of the ground state.

In our numerics we take $D = 100$ and the precision on the total energy d$E = 10^{-7}$ with a maximum number of sweeps equal to 6. For the points in which convergence is not reached within six sweeps $D$ is increased to 140 and additional sweeps are performed. We span the $(g,h)$ plane with a resolution of $0.02 \Delta$.

\subsection*{Phase boundaries}

Here we describe the criteria by which we numerically identify the phases shown in Fig.~\ref{fig_3}b of the main text. The border of the paramagnetic phase is immediately identified, since the phase consists of the points of the $g-h$ plane with magnetization per spin $M_z = -1/2$. Note that, given that $M_z$ commutes with $H$ and that the system is finite, the values of the total magnetization obtained numerically are integers, so that it is not necessary to define any tolerance for this order parameter. 

As explained in the main text, to identify the N\'eel ordered phase ``N" and the dimerized phase ``D" we examine the values of their characteristic order parameters $\Phi$ and $D_T$. In particular, we attribute a point to ``N" if $\Phi > 0.025$ and to ``D" if $M_z = 0$ and $D_T > 0.01$. To avoid contributions from the edges of the chain we evaluate the order parameters only in the 20 central atoms of the chain, even when the ''bulk" of the chain, i.e. the region where the expectation values of the observables are uniform with a certain period (2 for these two phases), is substantially more extended.

The lower border of the ``SMF" phase can be identified by comparing the magnetization profile obtained with the DMRG with that one predicted by $H^+$. One can indeed identify a value of the magnetic field $\bar{h}(g)$ for which $\partial M_z/\partial h$ increases for decreasing $h$, in qualitative disagreement with the behaviour of $M_z$ predicted by $H^+$. This happens at $M_z \approx -1/4$ and for $g \gtrsim 1.3\Delta$ the change in the magnetization curve is particularly sharp.

The borders of the trimerized phase ``T" are determined by taking those one of constant magnetization $M_z = -20/62 \sim -1/3$, and the condition that the expectation values of the observable are uniform with period 3 in the central region (20 atoms) of the chain, as in Fig.~\ref{fig_4}d of the main text. This happens for values of $g \gtrsim 1.3$, comparable with the coupling strength required for having dimerization.

\subsection*{Low-energy effective Hamiltonian}

In the case in which both the magnetic field $h$ and the coupling constant $g$ are much smaller than $\Delta$, one can obtain an effective Hamiltonian for the low-energy physics by applying Schrieffer-Wolff (SW) transformations  \cite{SW} on Hamiltonian \eqref{spin_chain_ham} of the main text. Before applying the transformation it is convenient to write the spin operators as fermions via a Jordan-Wigner transformation, and the pseudo-spin as Holstein-Primakoff bosons:
\begin{eqnarray}
\tilde{\sigma}^-_i&=&\sqrt{1-b^\dagger_ib_i}\,b_i,\\
\tilde{\sigma}^+_i&=&b^\dagger_i\,\sqrt{1-b^\dagger_ib_i},\\
\tilde{\sigma}^z_i&=&-1+2b^\dagger_ib_i.
\end{eqnarray}
This transformation is particularly convenient when, as in our case, $\braket{b^\dagger_ib_i} \ll 1$, since the square root operators can be expanded. Retaining up to second order in the bosonic operators we have 
\begin{multline}
\label{spin_chain_ham_hp_f}
H = (-\Delta+h)N + \sum_i\,2\Delta\,b^\dagger_ib_i - 2h\,c^\dagger_ic_i + \\
+ 2g\,\bigg\{(b^\dagger_i+b_i)(b^\dagger_{i+1}+b_{i+1}) - \chi(b^\dagger_i+b_i - b^\dagger_{i+1}-b_{i+1})\bigg\}\,\big(c^\dagger_ic_{i+1} + \text{h.c.}\big),
\end{multline}
where $\chi = \eta_a/(L\eta_0)$.
The SW transformation is based on the division of the Hilbert space into a low-energy subspace, that one with zero bosonic excitations, whose projector is $P_0$, and a high-energy one containing excitations whose projector is $Q_0$. For \eqref{spin_chain_ham_hp_f} it is convenient to apply the transformation separately for the term linear in bosonic operators $V_L = -2g\chi(b^\dagger_i+b_i - b^\dagger_{i+1}-b_{i+1})$ and the term quadratic $V_Q = 2g(b^\dagger_i+b_i)(b^\dagger_{i+1}+b_{i+1})$. We will show explicitly how the transformation acts on the first one. 

The effective Hamiltonian is given to the first order by 
\begin{equation}
H_L^{\mathrm{eff}} = \mathcal{D}(V_L) + \frac{1}{2}P_0[S, \mathcal{O}(V_L)]P_0,
\end{equation}
where the operators $\mathcal{D}(X) = P_0XP_0 + Q_0XQ_0$ and $\mathcal{O}(X) = P_0XQ_0 + Q_0XP_0$ take respectively the diagonal and the off-diagonal components of an operator (with respect to the low-energy and high-energy subspaces defined above), and 
\begin{equation}
S = \sum_{p,q} \frac{\bra{p}\mathcal{O}(V_L)\ket{q}}{E_p-E_q}\,\ket{p}\bra{q},
\end{equation}
with $\ket{p}, \ket{q}$ belonging to different subspaces. Clearly $\mathcal{D}(V_L) = 0$ and $\mathcal{O}(V_L) = V_L$ so that
\begin{equation}
\frac{1}{2}P_0[S, V_L]P_0 = -\frac{(2g\chi)^2}{2\Delta}\sum_{q,q'}\bra{q'}\sum_{i}\big(c^\dagger_ic_{i+1} - c^\dagger_{i-1}c_{i} + \text{h.c.}\big)^2\ket{q}\,\ket{q'}\bra{q}.
\end{equation}
Thus,
\begin{equation}
\label{H_eff_L}
H_L^{\mathrm{eff}} = - \frac{(2g\chi)^2}{2\Delta}\sum_{i}\big(c^\dagger_ic_{i+1} - c^\dagger_{i-1}c_{i} + \text{h.c.}\big)^2 = \frac{(2g\chi)^2}{2\Delta}\sum_{i}\,4c^\dagger_ic_i\big(c^\dagger_{i+1}c_{i+1} - 1\big) + \big(c^\dagger_{i-1}c_{i+1} + \text{h.c.}\big)\big(1-2c^\dagger_ic_i\big).
\end{equation}
Performing a similar transformation on the quadratic Hamiltonian $V_Q$ we obtain
\begin{equation}
\label{H_eff_Q}
H_Q^{\mathrm{eff}} = \frac{2g^2}{2\Delta}\sum_{i}\,c^\dagger_ic_i\big(c^\dagger_{i+1}c_{i+1} - 1\big).
\end{equation}
Combining the two transformations, which commute, we obtain 
\begin{equation}
H^{\mathrm{eff}} = \sum_{i}\,h\,\big(1-2c^\dagger_ic_i\big) + \frac{2g^2(1+4\chi^2)}{\Delta}\,c^\dagger_ic_i\big(c^\dagger_{i+1}c_{i+1} - 1\big) + \frac{2g^2\chi^2}{\Delta}\,\big(c^\dagger_{i-1}c_{i+1} + \text{h.c.}\big)\big(1-2c^\dagger_ic_i\big),
\end{equation}
which can be expressed by a JW transformation in terms of spin operators as Hamiltonian \eqref{Ham_eff} of the main text. 

\subsection*{Weak coupling phase diagram}

\begin{figure}[t]
	\setlength{\unitlength}{1cm}
	\begin{picture}(20,3)
	\put(2,0){\includegraphics[width=145mm,angle=0,clip]{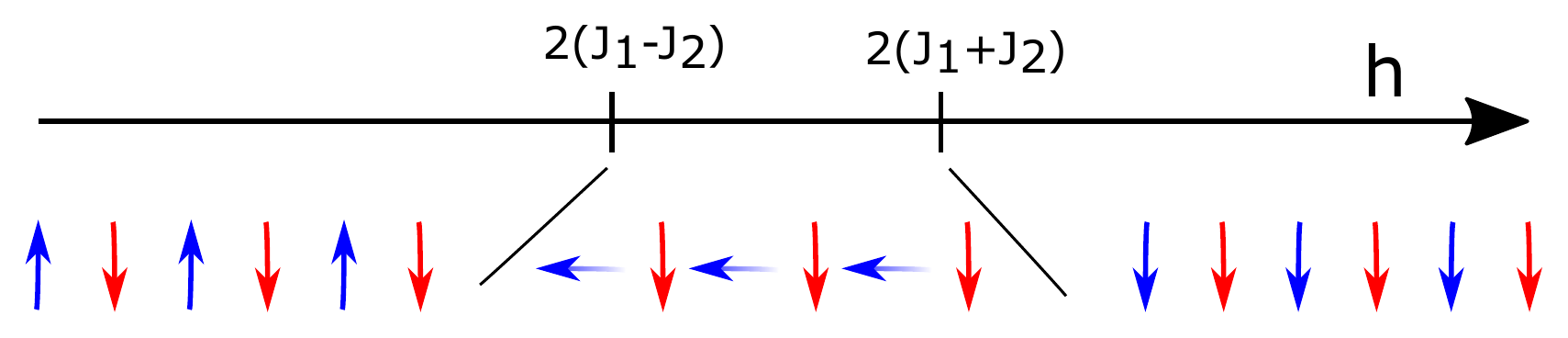}}
	\end{picture}
\caption{Schematic representation of the ground state spin configuration in the weak coupling regime. For $h < 2(J_1-J_2)$ the system is in a N\'eel ordered phase (``N''), while for $h > 2(J_1+J_2)$ the spin are all aligned (paramagnetic phase ``P''). For intermediate values of $h$ subchains A (in red in the figure) remains completely polarized, while subchain B ``melts'' under an effective XX model.}
\label{fig_1SI}
\end{figure}

The phase diagram of the weak coupling regime Hamiltonian \eqref{Ham_eff} can be easily studied considering the two subchains A and B of atoms at even and odd positions. For $h = 0$ the system has perfect antiferromagnetic order and all the spins of A (B) are in $\ket{\downarrow}$ ($\ket{\uparrow}$), given that $J_1 \gg J_2$. The intuition, confirmed by numerical results, tells us that the subchain A will not change if a positive magnetic field $h>0$ is turned on. Thus A effectively acts as an external field for subchain B, which then feels a total transversal field with amplitude $h_\mathrm{sub} = h - 2J_1$. The Hamiltonian for this subchain is thus
\begin{equation}
H^\textrm{eff}_B = \sum_{i}\,(h-2J_1)\sigma^z_i + 2J_2\,\big(\sigma^+_{i}\sigma^-_{i+1} + \sigma^-_{i}\sigma^+_{i+1}\big),
\end{equation}
which, being an XX model Hamiltonian can be solved immediately. It has phase transitions at $h = 2(J_1-J_2)$ and $h = 2(J_1+J_2)$ to paramagnetic phases, while between the two values the ground state is the XX gapless phase. Combining the states of the two subchains one gets the phase diagram pictured in Fig.~\ref{fig_1SI}.
The overall magnetization per atom of the chain can be easily calculated obtaining  
\begin{equation}
\label{M_Hp}
M_z = -1/2 + (1/2\pi)\arccos\bigg((h-2J_1)/2J_2\bigg).
\end{equation}

\subsection*{Large magnetization Hamiltonian}

Hamiltonian \eqref{spin_chain_ham} commutes with the total magnetization $M_z = (1/2)\sum_i \sigma^z_i$, but not with the operator $O = \sum_i \tilde{\sigma}^z_{i}\sigma^z_i$. The operator $O_i = \tilde{\sigma}^z_{i}\sigma^z_i$ has eigenvalues $\pm 1$, each with two-fold degeneracy. It is useful to decompose the Hamiltonian into a part $H'$ commuting with $O$ and a part $H''$ that does not. 

The part that commutes with $O$ consists of the non interacting part of \eqref{spin_chain_ham} as well as the $\tilde{\sigma}^x_i\tilde{\sigma}^x_{i+1}$ term in the interaction, and can be written as 
\begin{multline}
H' = \sum_i 1/2\big(\Delta + h\big)\big(\tilde{\sigma}^z_i + \sigma^z_i\big) + 1/2\big(\Delta - h\big)\big(\tilde{\sigma}^z_i - \sigma^z_i\big) + \\
+ 2g\big(\tilde{\sigma}^+_i\tilde{\sigma}^+_{i+1} + \tilde{\sigma}^+_i\tilde{\sigma}^-_{i+1} + \tilde{\sigma}^-_i\tilde{\sigma}^+_{i+1} + \tilde{\sigma}^-_i\tilde{\sigma}^-_{i+1}\big)\big(\sigma^+_i\sigma^-_{i+1} + \sigma^-_i\sigma^+_{i+1}\big).
\end{multline} 
It is convenient to introduce the following two sets of operators 
\begin{eqnarray}
\tau^{z}_i&=&(\tilde{\sigma}^z_i + \sigma^z_i)/2 \\
\tau^{+}_i&=&\tilde{\sigma}^+_i\sigma^+_i\\
\tau^{-}_i&=&\tilde{\sigma}^-_i\sigma^-_i
\end{eqnarray}
and
\begin{eqnarray}
\gamma^{z}_i&=&(-\tilde{\sigma}^z_i + \sigma^z_i)/2 \\
\gamma^{+}_i&=&\tilde{\sigma}^-_i\sigma^+_i\\
\gamma^{-}_i&=&\tilde{\sigma}^+_i\sigma^-_i
\end{eqnarray}
These two sets of operators obey the usual spin algebra, with operators from different sets commuting. The two sets of operators act on different subspaces of the local Hilbert space, corresponding to the two subspaces with eigenvalues $\pm 1$ of the operators $O_i$, i.e. $\mathcal{H}_i = \mathcal{H}^+_i \oplus \mathcal{H}^-_i$. We can thus write $H'$ as
\begin{equation}
H' = H^+ + H^- + H^{+-},
\end{equation}
where 
\begin{equation}
H^+ = \sum_i (\Delta+h)\tau^z_i + 2g(\tau^+_i\tau^-_{i+1} + \text{h.c.}),
\end{equation}
\begin{equation}
H^- = \sum_i (-\Delta+h)\gamma^z_i + 2g(\gamma^+_i\gamma^-_{i+1} + \text{h.c.})
\end{equation}
and 
\begin{equation}
H^{+-} = 2g\,\sum_i (\tau^+_i\gamma^-_{i+1} + \gamma^-_i\tau^+_{i+1} + \gamma^+_i\tau^-_{i+1} + \tau^-_i\gamma^+_{i+1}).
\end{equation}
$H''$ has a lengthy expression and will not be reproduced here.

As explained in the main text, when the system is nearly fully polarized (paramagnetic phase), with few spins flipped we have that $H''$ is much less relevant than $H'$. Note that the spins are prevalently in state $\ket{\Downarrow} = \ket{a,\downarrow}$, which belongs to the subspace described by $H^{+}$. One expects then that $H^{+}$ forms a good description of the system, with other terms in the Hamiltonian giving perturbative corrections. Note that $H^{+}$ connects the state $\ket{\Downarrow} = \ket{a,\downarrow}$ to $\ket{\Uparrow}= \ket{b,\uparrow}$, i.e. a simultaneous flip of the spin and band should occur.

\end{document}